\documentclass[10pt,aps,prl,twocolumn,superscriptaddress,floatfix,longbibliography]{revtex4-1}

\usepackage{graphicx}
\usepackage{xcolor}
\usepackage{amsmath}
\usepackage{amssymb}
\usepackage{amsfonts}
\usepackage{comment}
\usepackage{ulem}
\usepackage{braket}
\definecolor{scarred}{rgb}{0.75,0.0,0.0}
\usepackage{hyperref}
\hypersetup{
colorlinks=true,final=true,
        linkcolor=blue,
        citecolor=blue,
        filecolor=blue,
        urlcolor=blue,
}

\usepackage{mathtools}
\usepackage{amsbsy}

\begin{document}

\title{Nonlocal Correlation Effects in dc and Optical Conductivity of the Hubbard Model} 

\author{Nagamalleswararao Dasari}
\email{nagamalleswararao.d@gmail.com}
\affiliation{Institut f{\"u}r Theoretische Physik, Universit{\"a}t Hamburg, Notkestra{\ss}e   9 , 22607 Hamburg, Germany}

\author{Hugo U. R. Strand}
\affiliation{School of Science and Technology, Orebro University, SE-70182 Obrebo, Sweden}

\author{Martin Eckstein}
\affiliation{Institut f{\"u}r Theoretische Physik, Universit{\"a}t Hamburg, Notkestra{\ss}e   9 , 22607 Hamburg, Germany}
\affiliation{The Hamburg Centre for Ultrafast Imaging, Luruper Chaussee 149, 22761 Hamburg, Germany}

\author{Alexander I. Lichtenstein}
\affiliation{Institut f{\"u}r Theoretische Physik, Universit{\"a}t Hamburg, Notkestra{\ss}e   9 , 22607 Hamburg, Germany}
\affiliation{European X-Ray Free-Electron Laser Facility, Holzkoppel 4, 22869 Schenefeld, Germany}
\affiliation{The Hamburg Centre for Ultrafast Imaging, Luruper Chaussee 149, 22761 Hamburg, Germany}

\author{Evgeny A. Stepanov}
\affiliation{CPHT, CNRS, \'Ecole polytechnique, Institut Polytechnique de Paris, 91120 Palaiseau, France}
\affiliation{Coll\`ege de France, 11 place Marcelin Berthelot, 75005 Paris, France}

\begin{abstract}
Conductivity is one of the most direct probes of electronic systems, yet its theoretical description remains challenging in the presence of strong non-local correlations. 
In this Letter, we analyze the conductivity of the half-filled single-band Hubbard model and identify the role of spatial correlations across the Mott transition. 
We show that in the correlated metallic regime, an accurate description of the conductivity requires not only the correct spectral function but also the inclusion of complex multi-electron processes encoded in vertex corrections. 
The crossover to the Mott insulating regime is marked by a vanishing contribution of vertex corrections to the DC conductivity.
However, in the Mott insulating case vertex corrections remain significant for the optical conductivity.
\end{abstract}

\maketitle

Strong electronic correlations give rise to a variety of many-body phenomena, ranging from collective excitations (plasmons, magnons, Cooper pairs) to ordered phases including magnetic, superconducting, nematic, orbital-ordered, and charge-ordered states.
In correlated materials, these effects are experimentally investigated through various response functions, with transport measurements being among the simplest and most direct probes.
However, despite the apparent simplicity of these experiments, a reliable theoretical framework for accurately interpreting their results remains elusive.

Even the single-band Hubbard model, which is the minimal model that captures the interplay between electron kinetics on a lattice and on-site Coulomb repulsion, presents significant challenges for the theoretical calculation of transport properties.
A representative example is Ref.~\cite{doi:10.1126/science.aat4134}, which experimentally investigated the resistivity of the doped single-band Hubbard model in the strongly correlated regime using cold-atom simulations.
The authors have found that the dynamical mean-field theory (DMFT)~\cite{RevModPhys.68.13}, which is the state-of-the-art method for transport calculations of correlated materials, underestimates the resistivity compared to the experimental values at low temperatures and strongly overestimates it at high temperatures. 

Significant discrepancies have also been observed when comparing DMFT predictions with experimental data for real materials, such as the resistivity of various ruthenate compounds~\cite{PhysRevLett.116.256401, PhysRevMaterials.7.093801}.
As DMFT considers only local correlations, the observed discrepancies can naturally be attributed to the effects of missing spatial correlations.
These correlations can significantly influence transport properties by modifying the electronic spectral function and giving rise to complex multi-electron scattering processes, known as vertex corrections, which can strongly impact the conductivity. 
Incorporating non-local vertex corrections in state-of-the-art methods remains a major computational challenge. 
For this reason, the effect of spatial correlations on transport properties is largely unexplored, due to the lack of appropriate theoretical tools.

In this Letter, we systematically investigate the impact of non-local correlations on the conductivity of the single-band Hubbard model using the recently developed Dual $GW$ (${D\text{-}GW}$)~\cite{DGW} approach.
This method is ideally sited for this purpose, as it enables a consistent real-time description of local electronic correlations and long-range collective charge and spin fluctuations across weak- and strong-coupling regimes, offering direct access to single- and two-particle observables in real frequencies. By focusing on the region near the Mott transition, we find that the impact of non-local correlations on the conductivity differs between the correlated metallic and Mott insulating phases. 
We demonstrate that incorporating non-local correlations in both the electronic spectral function and vertex corrections is crucial for accurately describing the DC and optical conductivity in the metallic phase. 
The crossover between the metallic and Mott insulating phases is marked by a vanishing contribution of vertex corrections to the DC conductivity, although the DC conductivity itself remains finite at high temperatures. 
At the same time, non-local vertex corrections remain essential for describing the optical conductivity in the Mott insulating regime, despite the fact that this regime is dominated by local electronic correlations.

{\it Model and methods.}  The single-band Hubbard model with the on-site Coulomb repulsion $U$ and the hopping amplitude $t$ between the nearest-neighbor sites on a square lattice is described by the following Hamiltonian:
\begin{align}
\mathcal{H} = 
-t\sum_{\langle ij \rangle, \sigma} c^{\dagger}_{i\sigma}c^{\phantom{\dagger}}_{j\sigma}
+ U \sum_i n^{\phantom{\dagger}}_{i\uparrow} n^{\phantom{\dagger}}_{i\downarrow}, 
\label{eq:model}
\end{align}
where $c^{(\dagger)}_{i\sigma}$ annihilates (creates) an electron with spin ${\sigma \in \left\{ \uparrow,\downarrow \right\}}$ on the lattice site $i$, and ${n_{i\sigma} = c^\dagger_{i\sigma} c^{\phantom{\dagger}}_{i\sigma}}$ is the electronic density. 
The energy unit is set to half the bandwidth, ${D = 4t = 1}$, of the electronic dispersion. 
The electron Fermi surface (FS) of the half-filled model has a perfect nesting.
At weak coupling, this gives rise to the strong antiferromagnetic (AFM) fluctuations of itinerant electrons.
Upon lowering the temperature, these fluctuations lead to a momentum-selective opening of a gap in the electronic spectrum. The gap appears first at the antinodal [${\text{AN} = (\pi,0)}$] point of the FS, then gradually extends across the FS, eventually reaching the nodal [${\text{N} = (\frac{\pi}{2},\frac{\pi}{2})}$] point, as the system undergoes the N\'eel transition to the ordered AFM state.
At large interaction strengths, the system undergoes a transition into a Mott insulating state, which is driven by the local electronic correlations.
The weak and strong-coupling limits are connected by a rather broad crossover regime with coexisting itinerant fluctuations and local correlations~\cite{PhysRevLett.132.236504}. 

The identified correlation effects are expected to strongly influence transport properties of the system.
The study of electronic transport is usually conducted within linear response theory, where conductivity $\sigma_{\alpha\beta}(t,t')$ is defined as the linear coefficient relating the induced current $\langle j_{\alpha}(t)\rangle $ to an applied electromagnetic probe field $A_{\beta}(t')$, where ${\alpha, \beta \in \{x, y, z\}}$ label the direction in real space. 
The conductivity can be expressed as a sum of the bubble and renormalized terms that are explicitly detailed in the Supplemental Material (SM)~\cite{SM}. 
The bubble term accounts for the particle-hole excitation in the electronic spectral function through the convolution of the Green's functions.
The renormalized term additionally includes all possible scattering processes on collective electronic fluctuations, which are accounted for in the vertex correction.
In equilibrium, the conductivity \(\sigma_{\alpha \beta}(t,t')\) depends solely on the time difference, so the frequency-dependent conductivity $\sigma_{\alpha \beta}(\omega)$ can be obtained through a straightforward Fourier transform. 

Importantly, as demonstrated in the SM~\cite{SM}, the vertex correction originates from collective electronic fluctuations considered in the self-energy.
Moreover, the local part of the vertex is irrelevant for transport calculations, because it vanishes from the conductivity due to symmetry constraints~\cite{PhysRevLett.64.1990, RevModPhys.68.13, PhysRevLett.123.036601, SM}. 
This drastically simplifies the calculations in the DMFT framework, where the locality of the self-energy reduces the conductivity to the bubble term.
However, considering vertex corrections is sometimes crucial for a correct description of transport and optical properties of correlated systems~\cite{PhysRevB.104.245127, PhysRevLett.123.036601, acharya2023theory}. 
Incorporating the vertex corrections explicitly requires to solve the Bethe-Salpeter equation in momentum and time (or frequency) space, which is numerically expensive.
Extending beyond the local DMFT framework to consistently include both spatial correlations and non-local vertex corrections in conductivity calculations remains a major open problem. Recent diagrammatic methods have incorporated vertex corrections for optical conductivity perturbatively in weak-coupling approximations \cite{PhysRevB.104.245127, Kauch, Simard}. In contrast, cluster techniques include full vertex corrections but are limited to very high temperatures~\cite{PhysRevLett.123.036601}. A systematic study of vertex corrections to optical conductivity at low temperatures in the moderate to strong coupling regime has not yet been performed.

Alternatively, vertex corrections can be incorporated implicitly by introducing a time-dependent electric field and calculating conductivity as the derivative of the electronic current relative to this field~\cite{Tsuji,Martin2008,Shao,Dasari,PhysRevB.100.235117,Jaksa}. 
This approach closely mirrors experimental setups for transport measurements. 
At the same time, performing these calculations requires solving a time-dependent many-body problem, which is much more challenging than addressing the equilibrium one.
To resolve these issues, we have recently introduced a non-equilibrium Dual $GW$ (${D\text{-}GW}$) method~\cite{DGW}, which corresponds to the ``lighter'' version of the dual triply irreducible local expansion (\mbox{D-TRILEX}) approach~\cite{PhysRevB.100.205115, PhysRevB.103.245123, 10.21468/SciPostPhys.13.2.036} that allows for a real-time implementation of the method.
This theoretical framework provides a consistent diagrammatic extension of DMFT, which enables a simultaneous treatment of the spatial charge and spin fluctuations~\cite{PhysRevLett.127.207205, stepanov2021coexisting, PhysRevLett.129.096404, PhysRevResearch.5.L022016, vandelli2024doping, PhysRevLett.132.226501, stepanov2023charge, PhysRevLett.132.236504, PhysRevB.110.L161106, j6bj-gz7j, Ruthenates, Cuprates} thus improving upon the $GW$+DMFT method ~\cite{PhysRevLett.118.246402, PhysRevB.100.041111, PhysRevB.100.235117}, which is one of the most advanced theory for describing time evolution of correlated systems.
${D\text{-}GW}$ is therefore able to consistently account for the non-local vertex corrections in conductivity, particularly those arising from strong magnetic fluctuations, that are often the main source of instability in correlated systems.
Additionally, ${D\text{-}GW}$ is not restricted to a weak-coupling regime and thus provides a systematic framework to assess the relevance of vertex corrections across a broad range of model parameters.

\begin{figure}[t!]
\includegraphics[width=0.95\columnwidth]{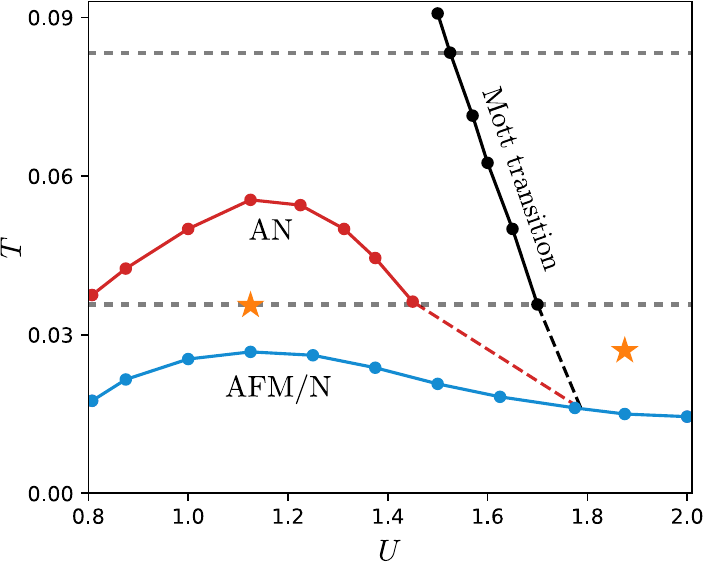}
\caption{Phase diagram of the half-filled single-band Hubbard model calculated using ${D\text{-}GW}$ in the $T$-$U$ plane. The blue curve represents the N\'eel transition and corresponds to the divergence of the spin susceptibility at the ${{\bf q} = (\pi, \pi)}$ point. The red curve indicates the gap opening at the AN point. 
The Mott transition (black curve) is characterized by a simultaneous gap opening at the N and AN points and is extrapolated to low temperatures by the dashed black line. The red dashed line is the extension of AN curve according to findings in Ref.~\cite{PhysRevLett.132.236504}. 
The orange stars indicate the two points at which the calculations are performed in Fig.~\ref{fig:fig2} and Fig.~\ref{fig:fig3}\,(a,\,b).
The dashed gray lines depict the temperatures at which the the optical conductivity scans are performed in Fig.~\ref{fig:fig3}\,(c).}  
\label{fig:phase}
\end{figure}

The phase diagram of the model~\eqref{eq:model} obtained from ${D\text{-}GW}$ is shown in Fig.~\ref{fig:phase} and is similar to the one of Ref.~\cite{PhysRevLett.132.236504}. 
The AFM  transition is represented by a blue curve.
Above this curve, in the paramagnetic phase the red curve depicts the opening of the gap at the AN point.
The crossover to the Mott insulating state is shown by the black line and is characterized by a simultaneous opening of the gap at all ${\bf k}$-points of the FS~\cite{PhysRevLett.132.236504}.
Details of the numerical calculations are provided in the SM~\cite{SM}.

\begin{figure}[t!]
\includegraphics[width=0.95\columnwidth]{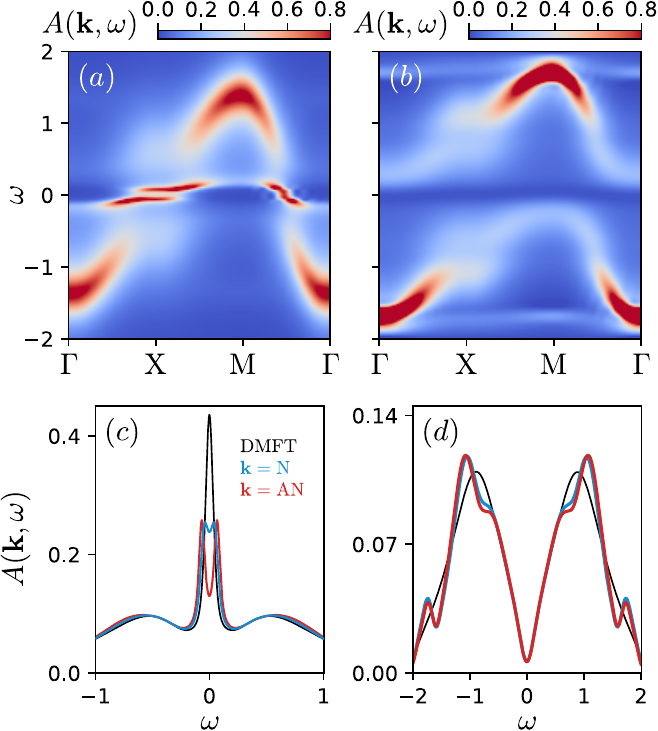}
\caption{The momentum-resolved spectral function ${A({\bf k},\omega)}$ calculated close to the N\'eel transition (orange stars in Fig.~\ref{fig:phase}) using ${D\text{-}GW}$ along the high-symmetry path of the Brillouin zone (${\Gamma=(0,0)}$, ${\text{X}=\text{AN}=(\pi,0)}$, ${\text{M}=(\pi,\pi)}$) in the metallic (${U=1.125}$, ${T = 0.036}$, panel (a)) and Mott insulating (${U=1.875}$, ${T = 0.027}$, panel (b)) phases. Panels (c) and (d) show the corresponding spectral functions at the $N$ (blue) and $AN$ (red) points in comparison with DMFT (black). 
\label{fig:fig2}}
\end{figure}

{\it Results.} Optical conductivity is primarily governed by single-particle excitations, making an accurate electronic spectral function essential for its calculation.
Strong magnetic fluctuations can extend to rather high temperatures above the AFM transition (see SM~\cite{SM} for details) and can drastically modify the electronic spectral function.
Thus, in the metallic phase, they can result in the opening of a gap at some part of the Fermi surface, as illustrated in Fig.~\ref{fig:fig2}\,(a) (see also Ref.~\cite{Geng}). 
The comparison of spectra at the N and AN points with DMFT in Fig.~\ref{fig:fig2}\,(c) shows that in ${D\text{-}GW}$ this gap opening is momentum-dependent.
In turn, DMFT does not account for spatial fluctuations, resulting in a clear quasi-particle peak at the Fermi level. 
In the Mott insulating case, strong magnetic fluctuations affect the electronic spectrum mainly at high energies, resulting in the splitting of Hubbard bands at the N and AN points, as illustrated in Fig.~\ref{fig:fig2}\,(b).
A three-peak structure of the Hubbard band obtained on both sides of the Mott-gap using ${D\text{-}GW}$ signifies the role of spatial spin fluctuations in comparison with DMFT in Fig.~\ref{fig:fig2}\,(d). 
We note that a similar structure of Hubbard bands is found in variational Monte-Carlo calculations~\cite{PhysRevX.10.041023, singh2022unconventional} for the same model~\eqref{eq:model}, and also in \mbox{D-TRILEX} calculations for the two-orbital model in the presence of strong magnetic fluctuations~\cite{PhysRevLett.129.096404}.

\begin{figure}[t!]
\includegraphics[width=0.95\columnwidth]{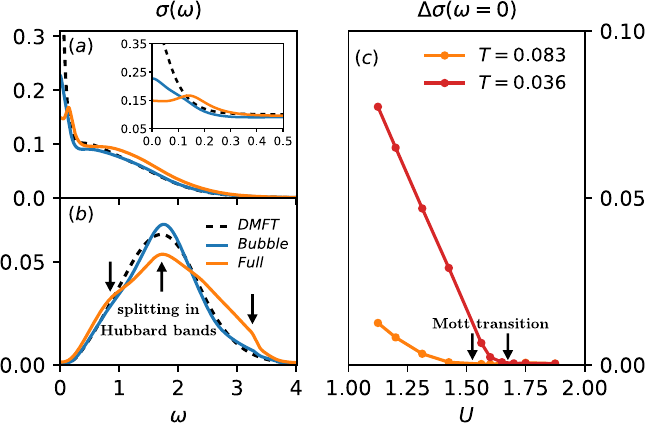}
\caption{The optical conductivity ${\sigma(\omega)}$ calculated close to the N\'eel transition (orange stars in Fig.~\ref{fig:phase}) in the metallic (${U = 1.125}$, ${T = 0.036}$, panel (a)) and Mott insulating (${U = 1.875}$,  ${T = 0.027}$, panel (b)) phases. 
The results are obtained using DMFT (dashed black), ``bubble'' ${D\text{-}GW}$ (blue), and ``full'' ${D\text{-}GW}$ (orange). 
The inset in (a) shows the low-frequency behavior of the conductivity. The arrows in panel (b) indicate the peaks and kinks originating from the optical transitions between the splittings in Hubbard bands due to strong magnetic fluctuations seen
in Fig.~\ref{fig:fig2}\,(d).
Panel (c) shows the difference in the DC conductivity $\Delta{\sigma(\omega=0)}$ between the bubble and full results. The scans are performed for ${T=0.083}$ and ${T=0.036}$ depicted in Fig.~\ref{fig:phase} by the dashed gray lines. 
The arrows in panel (c) indicate critical interactions ${U = 1.525}$ (${T=0.083}$) and ${U = 1.7}$ (${T=0.036}$) for the Mott transition shown in black dots in Fig.~\ref{fig:phase}.}
\label{fig:fig3}
\end{figure}

One can naively expect that if a method accurately captures the single-particle spectral features, the bubble approximation for the current–current correlation function would be sufficient to describe the optical conductivity qualitatively.
To demonstrate that this assumption does not hold, we calculate the optical conductivity in both the metallic and Mott insulating regimes, and compare the results obtained from DMFT, ${D\text{-}GW}$ within the ``bubble'' approximation, and the ``full'' ${D\text{-}GW}$ calculation that includes vertex corrections.
This comparison allows us to disentangle the different impacts of correlations on the conductivity, as DMFT accounts only for local correlations, the bubble ${D\text{-}GW}$ approximation additionally includes the effect of non-local correlations, but only on the electronic spectral function, and the full ${D\text{-}GW}$ scheme further incorporates non-local correlations via the vertex corrections.
The results obtained for the metallic case are shown in Fig~\ref{fig:fig3}\,(a). 
In DMFT, we observe a finite spectral weight at low energy, corresponding to the Drude peak, followed by incoherent excitations at higher energies. 
In the bubble ${D\text{-}GW}$ approximation, we find that the Drude weight, corresponding to the DC conductivity, is significantly suppressed, while the optical (high-energy) part of the conductivity remains essentially unchanged compared to DMFT. 
In the full ${D\text{-}GW}$ calculation, the DC conductivity is further suppressed relative to the bubble approximation, and an additional peak appears at ${\omega \simeq 0.15}$.
This peak reflects the formation of a gap in the electronic spectral function (Fig.~\ref{fig:fig2}\,(c)) driven by strong magnetic fluctuations. 
Remarkably, this peak is not present in the bubble approximation, although the later accounts for the correct electronic spectral function through the Green's functions.
The high-energy part of the conductivity is also noticeably modified upon considering vertex corrections in the metallic case. 

The transition from metal to Mott insulator is driven by local electronic correlations. 
Therefore, one may expect that vertex corrections to the conductivity are negligible, as the locality of the self-energy leads to local vertex corrections that, by symmetry, drop out of the expression for the current–current correlation function.
In this case, the bubble approximation should provide an accurate description of the optical conductivity in the Mott insulating phase. 
However, it is evident from the splitting of Hubbard bands (Fig.~\ref{fig:fig2}\,b,\,d), that non-local spin fluctuations can strongly affect the electronic spectral function even in the Mott phase. 
In order to illustrate how spin fluctuations influence conductivity, in Fig.~\ref{fig:fig3}\,(b) we compare results obtained using DMFT (black dashed curve), bubble ${D\text{-}GW}$ approximation (blue curve), and full ${D\text{-}GW}$ scheme (orange curve).
Apart from low energy, the optical conductivity from the bubble approximation differ significantly from the full calculation.
The full ${D\text{-}GW}$ results show peaks and kinks at three marked energies (indicated by black arrows), resembling those of the single-particle spectral function illustrated in Fig.~\ref{fig:fig2}\,(d).
DMFT, instead, shows only one peak corresponding to resonant excitations between Hubbard bands.
Surprisingly, the bubble approximation, although based on the correct form of the electronic spectral function, does not show a clear three-peak form of the conductivity observed in the full ${D\text{-}GW}$ calculation. 
It only reveals a ``resonant'' transition between the Hubbard bands, and a slight change in the curvature at frequencies, where the full calculation finds two additional peaks. 
We note that experimental measurements on iridate compounds have revealed a similarly peaked structure in the optical conductivity~\cite{PhysRevLett.101.076402, PhysRevB.80.195110, seo2017infrared, 10.1063/1.4870049}. 
In the theoretical work~\cite{Cassol} these peaks were related to the splitting of Hubbard bands due to antiferromagnetic fluctuations that are strong in these materials~\cite{Lenz_2019}. 
However, the bubble approximation for conductivity, that accounts for the band splitting, captured only part of this feature by producing a less intense precursor of the experimentally measured peaks~\cite{Cassol}.
Therefore, accurately reproducing the optical conductivity in the Mott phase still necessary requires considering non-local vertex corrections related to electronic scattering on spatial magnetic fluctuations. 
Additionally, the conductivity in both metallic and Mott regimes can be affected by the non-local Coulomb interaction. Its role is explicitly discussed in the SM~\cite{SM}.

At the same time, we observe that the low-frequency part of the conductivity in the Mott phase does not vary significantly across different approximations. 
The DC conductivity, ${\sigma(\omega = 0)}$, is primarily governed by the electronic spectral weight at the Fermi energy. 
Accordingly, the opening of the Mott gap should be directly reflected in the DC conductivity. 
Since this gap opening is momentum-independent (i.e., local), one can expect the role of vertex corrections in the DC conductivity to vanish upon entering the Mott insulating phase.
In Fig.~\ref{fig:fig3}\,(c), we show the difference in DC conductivity, ${\Delta\sigma(\omega = 0)}$, between the bubble and full ${D\text{-}GW}$ results for two different temperatures, plotted as a function of interaction strength $U$. 
We find that ${\Delta\sigma(\omega = 0)}$ decreases linearly with increasing $U$ in the metallic phase and vanishes at the critical interaction strength indicated by the black arrows, where the Mott transition occurs.
We note that at sufficiently high temperatures, the first-order Mott transition evolves into a smooth crossover, during which the quasiparticle peak at the Fermi energy transforms into a minimum, while the electronic density at the Fermi level remains finite due to thermal fluctuations. 
Consequently, the DC conductivity in the finite-temperature Mott regime is not zero, as illustrated in the SM~\cite{SM}.
We have confirmed that the critical interaction values for the Mott transition, as identified via the DC conductivity, are consistent with those obtained from the electronic spectral function.

\begin{figure}[t!]
\includegraphics[width=0.95\columnwidth]{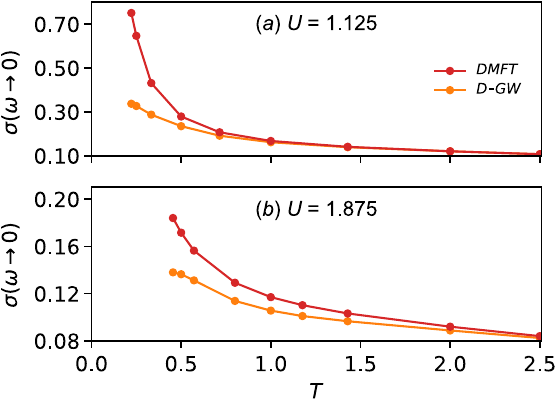}
\caption{The DC conductivity as a function of temperature calculated for ${n=0.85}$ at ${U=1.125}$ (a) and ${U=1.875}$ (b) using the DMFT (red) and ``full'' {\it D-GW} (orange) methods.}
\label{fig:fig4}
\end{figure}

Finally, to set the stage for investigation of anomalous transport in realistic materials, we perform DMFT and full {\it D-GW} calculations at 15\% hole doping.
The resulting temperature-dependent DC conductivity is shown in Fig.~\ref{fig:fig4}.
We find that, while negligible at high temperatures, the contribution of non-local correlations to the DC conductivity increases substantially upon lowering the temperature in both the intermediate (a) and strong (b) coupling regimes.
In both cases, incorporating spatial magnetic fluctuations within the {\it D-GW} framework leads to a suppression of the DC conductivity, consistent with Ref.~\cite{doi:10.1126/science.aat4134}, where DMFT was also found to underestimate the experimentally measured resistivity in the low-temperature regime.
This improved description of the DC conductivity suggests that the {\it D-GW} framework provides a promising route toward tackling more challenging transport problems, including the ``strange metal'' behavior observed in cuprates, iron pnictides, ruthenates, and cobaltates~\cite{keimer2015quantum, doi:10.1126/science.abh4273}.

{\it Conclusions.} 
In conclusion, we investigated the influence of strong non-local spin fluctuations on the optical conductivity of the half-filled Hubbard model. 
We found that neglecting vertex corrections associated with spin fluctuations is justified only for the DC conductivity in the Mott insulating phase, where the gap opening is momentum-independent. 
In contrast, in the metallic phase, where the gap opening is momentum-dependent, these vertices are crucial for determining the DC conductivity.
Moreover, vertex corrections are essential for capturing finite-frequency peaks in the conductivity, which arise from momentum-dependent renormalization of the electronic spectral function driven by strong magnetic fluctuations, such as the splitting of the quasiparticle peak at the Fermi level in the metallic phase or the splitting of the Hubbard bands in the Mott insulating phase.

\begin{acknowledgments}
The authors thank Maria Chatzieleftheriou and Francesco Cassol for fruitful discussions. 
N.D., H.U.R.S., and A.I.L. acknowledge support from the European Research Council via Synergy Grant No.\ 854843 (the FASTCORR project). 
H.U.R.S.\ acknowledges financial support from the Swedish Research Council (Vetenskapsrådet, VR) grant number 2024-04652. M.E. and A.I.L. acknowledge support from the Deutsche Forschungs-
gemeinschaft through the research unit QUAST, FOR 5249,
Project ID No. 449872909. 
E.A.S. acknowledges support from TGCC-GENCI through the AD010901393R1 project, LabEx PALM Paris-Saclay through the CEBULI project, and CNRS through the Physique Tremplin project UFEX. This research was supported in part through the EuXFEL (Maxwell) computational resources operated at Deutsches Elektronen-Synchrotron DESY, Hamburg, Germany.
\end{acknowledgments}

\bibliography{Ref}

\begin{thebibliography}{48}%
\makeatletter
\providecommand \@ifxundefined [1]{%
 \@ifx{#1\undefined}
}%
\providecommand \@ifnum [1]{%
 \ifnum #1\expandafter \@firstoftwo
 \else \expandafter \@secondoftwo
 \fi
}%
\providecommand \@ifx [1]{%
 \ifx #1\expandafter \@firstoftwo
 \else \expandafter \@secondoftwo
 \fi
}%
\providecommand \natexlab [1]{#1}%
\providecommand \enquote  [1]{``#1''}%
\providecommand \bibnamefont  [1]{#1}%
\providecommand \bibfnamefont [1]{#1}%
\providecommand \citenamefont [1]{#1}%
\providecommand \href@noop [0]{\@secondoftwo}%
\providecommand \href [0]{\begingroup \@sanitize@url \@href}%
\providecommand \@href[1]{\@@startlink{#1}\@@href}%
\providecommand \@@href[1]{\endgroup#1\@@endlink}%
\providecommand \@sanitize@url [0]{\catcode `\\12\catcode `\$12\catcode
  `\&12\catcode `\#12\catcode `\^12\catcode `\_12\catcode `\%12\relax}%
\providecommand \@@startlink[1]{}%
\providecommand \@@endlink[0]{}%
\providecommand \url  [0]{\begingroup\@sanitize@url \@url }%
\providecommand \@url [1]{\endgroup\@href {#1}{\urlprefix }}%
\providecommand \urlprefix  [0]{URL }%
\providecommand \Eprint [0]{\href }%
\providecommand \doibase [0]{http://dx.doi.org/}%
\providecommand \selectlanguage [0]{\@gobble}%
\providecommand \bibinfo  [0]{\@secondoftwo}%
\providecommand \bibfield  [0]{\@secondoftwo}%
\providecommand \translation [1]{[#1]}%
\providecommand \BibitemOpen [0]{}%
\providecommand \bibitemStop [0]{}%
\providecommand \bibitemNoStop [0]{.\EOS\space}%
\providecommand \EOS [0]{\spacefactor3000\relax}%
\providecommand \BibitemShut  [1]{\csname bibitem#1\endcsname}%
\let\auto@bib@innerbib\@empty
\bibitem [{\citenamefont {Brown}\ \emph {et~al.}(2019)\citenamefont {Brown},
  \citenamefont {Mitra}, \citenamefont {Guardado-Sanchez}, \citenamefont
  {Nourafkan}, \citenamefont {Reymbaut}, \citenamefont {Hébert}, \citenamefont
  {Bergeron}, \citenamefont {Tremblay}, \citenamefont {Kokalj}, \citenamefont
  {Huse}, \citenamefont {Schauß},\ and\ \citenamefont
  {Bakr}}]{doi:10.1126/science.aat4134}%
  \BibitemOpen
  \bibfield  {author} {\bibinfo {author} {\bibfnamefont {Peter~T.}\
  \bibnamefont {Brown}}, \bibinfo {author} {\bibfnamefont {Debayan}\
  \bibnamefont {Mitra}}, \bibinfo {author} {\bibfnamefont {Elmer}\ \bibnamefont
  {Guardado-Sanchez}}, \bibinfo {author} {\bibfnamefont {Reza}\ \bibnamefont
  {Nourafkan}}, \bibinfo {author} {\bibfnamefont {Alexis}\ \bibnamefont
  {Reymbaut}}, \bibinfo {author} {\bibfnamefont {Charles-David}\ \bibnamefont
  {Hébert}}, \bibinfo {author} {\bibfnamefont {Simon}\ \bibnamefont
  {Bergeron}}, \bibinfo {author} {\bibfnamefont {A.-M.~S.}\ \bibnamefont
  {Tremblay}}, \bibinfo {author} {\bibfnamefont {Jure}\ \bibnamefont {Kokalj}},
  \bibinfo {author} {\bibfnamefont {David~A.}\ \bibnamefont {Huse}}, \bibinfo
  {author} {\bibfnamefont {Peter}\ \bibnamefont {Schauß}}, \ and\ \bibinfo
  {author} {\bibfnamefont {Waseem~S.}\ \bibnamefont {Bakr}},\ }\bibfield
  {title} {\enquote {\bibinfo {title} {{Bad metallic transport in a cold atom
  Fermi-Hubbard system}},}\ }\href {\doibase 10.1126/science.aat4134}
  {\bibfield  {journal} {\bibinfo  {journal} {Science}\ }\textbf {\bibinfo
  {volume} {363}},\ \bibinfo {pages} {379--382} (\bibinfo {year}
  {2019})}\BibitemShut {NoStop}%
\bibitem [{\citenamefont {Georges}\ \emph {et~al.}(1996)\citenamefont
  {Georges}, \citenamefont {Kotliar}, \citenamefont {Krauth},\ and\
  \citenamefont {Rozenberg}}]{RevModPhys.68.13}%
  \BibitemOpen
  \bibfield  {author} {\bibinfo {author} {\bibfnamefont {Antoine}\ \bibnamefont
  {Georges}}, \bibinfo {author} {\bibfnamefont {Gabriel}\ \bibnamefont
  {Kotliar}}, \bibinfo {author} {\bibfnamefont {Werner}\ \bibnamefont
  {Krauth}}, \ and\ \bibinfo {author} {\bibfnamefont {Marcelo~J.}\ \bibnamefont
  {Rozenberg}},\ }\bibfield  {title} {\enquote {\bibinfo {title} {{Dynamical
  mean-field theory of strongly correlated fermion systems and the limit of
  infinite dimensions}},}\ }\href {\doibase 10.1103/RevModPhys.68.13}
  {\bibfield  {journal} {\bibinfo  {journal} {Rev. Mod. Phys.}\ }\textbf
  {\bibinfo {volume} {68}},\ \bibinfo {pages} {13--125} (\bibinfo {year}
  {1996})}\BibitemShut {NoStop}%
\bibitem [{\citenamefont {Deng}\ \emph {et~al.}(2016)\citenamefont {Deng},
  \citenamefont {Haule},\ and\ \citenamefont
  {Kotliar}}]{PhysRevLett.116.256401}%
  \BibitemOpen
  \bibfield  {author} {\bibinfo {author} {\bibfnamefont {Xiaoyu}\ \bibnamefont
  {Deng}}, \bibinfo {author} {\bibfnamefont {Kristjan}\ \bibnamefont {Haule}},
  \ and\ \bibinfo {author} {\bibfnamefont {Gabriel}\ \bibnamefont {Kotliar}},\
  }\bibfield  {title} {\enquote {\bibinfo {title} {{Transport Properties of
  Metallic Ruthenates: A $\mathrm{DFT}+\mathrm{DMFT}$ Investigation}},}\ }\href
  {\doibase 10.1103/PhysRevLett.116.256401} {\bibfield  {journal} {\bibinfo
  {journal} {Phys. Rev. Lett.}\ }\textbf {\bibinfo {volume} {116}},\ \bibinfo
  {pages} {256401} (\bibinfo {year} {2016})}\BibitemShut {NoStop}%
\bibitem [{\citenamefont {Abramovitch}\ \emph {et~al.}(2023)\citenamefont
  {Abramovitch}, \citenamefont {Zhou}, \citenamefont {Mravlje}, \citenamefont
  {Georges},\ and\ \citenamefont {Bernardi}}]{PhysRevMaterials.7.093801}%
  \BibitemOpen
  \bibfield  {author} {\bibinfo {author} {\bibfnamefont {David~J.}\
  \bibnamefont {Abramovitch}}, \bibinfo {author} {\bibfnamefont {Jin-Jian}\
  \bibnamefont {Zhou}}, \bibinfo {author} {\bibfnamefont {Jernej}\ \bibnamefont
  {Mravlje}}, \bibinfo {author} {\bibfnamefont {Antoine}\ \bibnamefont
  {Georges}}, \ and\ \bibinfo {author} {\bibfnamefont {Marco}\ \bibnamefont
  {Bernardi}},\ }\bibfield  {title} {\enquote {\bibinfo {title} {{Combining
  electron-phonon and dynamical mean-field theory calculations of correlated
  materials: Transport in the correlated metal
  ${\mathrm{Sr}}_{2}{\mathrm{RuO}}_{4}$}},}\ }\href {\doibase
  10.1103/PhysRevMaterials.7.093801} {\bibfield  {journal} {\bibinfo  {journal}
  {Phys. Rev. Mater.}\ }\textbf {\bibinfo {volume} {7}},\ \bibinfo {pages}
  {093801} (\bibinfo {year} {2023})}\BibitemShut {NoStop}%
\bibitem [{\citenamefont {Dasari}\ \emph {et~al.}(2025)\citenamefont {Dasari},
  \citenamefont {Strand}, \citenamefont {Eckstein}, \citenamefont
  {Lichtenstein},\ and\ \citenamefont {Stepanov}}]{DGW}%
  \BibitemOpen
  \bibfield  {author} {\bibinfo {author} {\bibfnamefont {Nagamalleswararao}\
  \bibnamefont {Dasari}}, \bibinfo {author} {\bibfnamefont {Hugo U.~R.}\
  \bibnamefont {Strand}}, \bibinfo {author} {\bibfnamefont {Martin}\
  \bibnamefont {Eckstein}}, \bibinfo {author} {\bibfnamefont {Alexander~I.}\
  \bibnamefont {Lichtenstein}}, \ and\ \bibinfo {author} {\bibfnamefont
  {Evgeny~A.}\ \bibnamefont {Stepanov}},\ }\bibfield  {title} {\enquote
  {\bibinfo {title} {{Electron-magnon dynamics triggered by an ultrashort laser
  pulse: A real-time dual $GW$ study}},}\ }\href {\doibase 10.1103/vglv-2rmv}
  {\bibfield  {journal} {\bibinfo  {journal} {Phys. Rev. B}\ }\textbf {\bibinfo
  {volume} {111}},\ \bibinfo {pages} {235129} (\bibinfo {year}
  {2025})}\BibitemShut {NoStop}%
\bibitem [{\citenamefont {Chatzieleftheriou}\ \emph {et~al.}(2024)\citenamefont
  {Chatzieleftheriou}, \citenamefont {Biermann},\ and\ \citenamefont
  {Stepanov}}]{PhysRevLett.132.236504}%
  \BibitemOpen
  \bibfield  {author} {\bibinfo {author} {\bibfnamefont {Maria}\ \bibnamefont
  {Chatzieleftheriou}}, \bibinfo {author} {\bibfnamefont {Silke}\ \bibnamefont
  {Biermann}}, \ and\ \bibinfo {author} {\bibfnamefont {Evgeny~A.}\
  \bibnamefont {Stepanov}},\ }\bibfield  {title} {\enquote {\bibinfo {title}
  {{Local and Nonlocal Electronic Correlations at the Metal-Insulator
  Transition in the Two-Dimensional Hubbard Model}},}\ }\href {\doibase
  10.1103/PhysRevLett.132.236504} {\bibfield  {journal} {\bibinfo  {journal}
  {Phys. Rev. Lett.}\ }\textbf {\bibinfo {volume} {132}},\ \bibinfo {pages}
  {236504} (\bibinfo {year} {2024})}\BibitemShut {NoStop}%
\bibitem [{SM()}]{SM}%
  \BibitemOpen
  \bibinfo {note} {See Supplemental Material [url] for the explicit
  diagrammatic form of conductivity, the relation between the self-energy and
  vertex corrections in conductivity, the fixed temperature scans of
  conductivity at different interactions and details of the numerical
  calculations. The Supplemental Material includes
  Refs.~\cite{PhysRevB.82.115115, SCHULER2020107484}.}\BibitemShut {Stop}%
\bibitem [{\citenamefont {Khurana}(1990)}]{PhysRevLett.64.1990}%
  \BibitemOpen
  \bibfield  {author} {\bibinfo {author} {\bibfnamefont {Anil}\ \bibnamefont
  {Khurana}},\ }\bibfield  {title} {\enquote {\bibinfo {title} {{Electrical
  conductivity in the infinite-dimensional Hubbard model}},}\ }\href {\doibase
  10.1103/PhysRevLett.64.1990} {\bibfield  {journal} {\bibinfo  {journal}
  {Phys. Rev. Lett.}\ }\textbf {\bibinfo {volume} {64}},\ \bibinfo {pages}
  {1990--1990} (\bibinfo {year} {1990})}\BibitemShut {NoStop}%
\bibitem [{\citenamefont {Vu\ifmmode \check{c}\else \v{c}\fi{}i\ifmmode
  \check{c}\else \v{c}\fi{}evi\ifmmode~\acute{c}\else \'{c}\fi{}}\ \emph
  {et~al.}(2019)\citenamefont {Vu\ifmmode \check{c}\else \v{c}\fi{}i\ifmmode
  \check{c}\else \v{c}\fi{}evi\ifmmode~\acute{c}\else \'{c}\fi{}},
  \citenamefont {Kokalj}, \citenamefont {\ifmmode~\check{Z}\else
  \v{Z}\fi{}itko}, \citenamefont {Wentzell}, \citenamefont
  {Tanaskovi\ifmmode~\acute{c}\else \'{c}\fi{}},\ and\ \citenamefont
  {Mravlje}}]{PhysRevLett.123.036601}%
  \BibitemOpen
  \bibfield  {author} {\bibinfo {author} {\bibfnamefont {J.}~\bibnamefont
  {Vu\ifmmode \check{c}\else \v{c}\fi{}i\ifmmode \check{c}\else
  \v{c}\fi{}evi\ifmmode~\acute{c}\else \'{c}\fi{}}}, \bibinfo {author}
  {\bibfnamefont {J.}~\bibnamefont {Kokalj}}, \bibinfo {author} {\bibfnamefont
  {R.}~\bibnamefont {\ifmmode~\check{Z}\else \v{Z}\fi{}itko}}, \bibinfo
  {author} {\bibfnamefont {N.}~\bibnamefont {Wentzell}}, \bibinfo {author}
  {\bibfnamefont {D.}~\bibnamefont {Tanaskovi\ifmmode~\acute{c}\else
  \'{c}\fi{}}}, \ and\ \bibinfo {author} {\bibfnamefont {J.}~\bibnamefont
  {Mravlje}},\ }\bibfield  {title} {\enquote {\bibinfo {title} {{Conductivity
  in the Square Lattice Hubbard Model at High Temperatures: Importance of
  Vertex Corrections}},}\ }\href {\doibase 10.1103/PhysRevLett.123.036601}
  {\bibfield  {journal} {\bibinfo  {journal} {Phys. Rev. Lett.}\ }\textbf
  {\bibinfo {volume} {123}},\ \bibinfo {pages} {036601} (\bibinfo {year}
  {2019})}\BibitemShut {NoStop}%
\bibitem [{\citenamefont {Simard}\ \emph
  {et~al.}(2021{\natexlab{a}})\citenamefont {Simard}, \citenamefont
  {Eckstein},\ and\ \citenamefont {Werner}}]{PhysRevB.104.245127}%
  \BibitemOpen
  \bibfield  {author} {\bibinfo {author} {\bibfnamefont {Olivier}\ \bibnamefont
  {Simard}}, \bibinfo {author} {\bibfnamefont {Martin}\ \bibnamefont
  {Eckstein}}, \ and\ \bibinfo {author} {\bibfnamefont {Philipp}\ \bibnamefont
  {Werner}},\ }\bibfield  {title} {\enquote {\bibinfo {title} {{Nonequilibrium
  evolution of the optical conductivity of the weakly interacting Hubbard
  model: Drude response and $\ensuremath{\pi}$-ton type vertex corrections}},}\
  }\href {\doibase 10.1103/PhysRevB.104.245127} {\bibfield  {journal} {\bibinfo
   {journal} {Phys. Rev. B}\ }\textbf {\bibinfo {volume} {104}},\ \bibinfo
  {pages} {245127} (\bibinfo {year} {2021}{\natexlab{a}})}\BibitemShut
  {NoStop}%
\bibitem [{\citenamefont {Acharya}\ \emph {et~al.}(2023)\citenamefont
  {Acharya}, \citenamefont {Pashov}, \citenamefont {Weber}, \citenamefont {van
  Schilfgaarde}, \citenamefont {Lichtenstein},\ and\ \citenamefont
  {Katsnelson}}]{acharya2023theory}%
  \BibitemOpen
  \bibfield  {author} {\bibinfo {author} {\bibfnamefont {Swagata}\ \bibnamefont
  {Acharya}}, \bibinfo {author} {\bibfnamefont {Dimitar}\ \bibnamefont
  {Pashov}}, \bibinfo {author} {\bibfnamefont {Cedric}\ \bibnamefont {Weber}},
  \bibinfo {author} {\bibfnamefont {Mark}\ \bibnamefont {van Schilfgaarde}},
  \bibinfo {author} {\bibfnamefont {Alexander~I}\ \bibnamefont {Lichtenstein}},
  \ and\ \bibinfo {author} {\bibfnamefont {Mikhail~I}\ \bibnamefont
  {Katsnelson}},\ }\bibfield  {title} {\enquote {\bibinfo {title} {{A theory
  for colors of strongly correlated electronic systems}},}\ }\href {\doibase
  10.1038/s41467-023-41314-6} {\bibfield  {journal} {\bibinfo  {journal} {Nat.
  Commun.}\ }\textbf {\bibinfo {volume} {14}},\ \bibinfo {pages} {5565}
  (\bibinfo {year} {2023})}\BibitemShut {NoStop}%
\bibitem [{\citenamefont {Kauch}\ \emph {et~al.}(2020)\citenamefont {Kauch},
  \citenamefont {Pudleiner}, \citenamefont {Astleithner}, \citenamefont
  {Thunstr\"om}, \citenamefont {Ribic},\ and\ \citenamefont {Held}}]{Kauch}%
  \BibitemOpen
  \bibfield  {author} {\bibinfo {author} {\bibfnamefont {A.}~\bibnamefont
  {Kauch}}, \bibinfo {author} {\bibfnamefont {P.}~\bibnamefont {Pudleiner}},
  \bibinfo {author} {\bibfnamefont {K.}~\bibnamefont {Astleithner}}, \bibinfo
  {author} {\bibfnamefont {P.}~\bibnamefont {Thunstr\"om}}, \bibinfo {author}
  {\bibfnamefont {T.}~\bibnamefont {Ribic}}, \ and\ \bibinfo {author}
  {\bibfnamefont {K.}~\bibnamefont {Held}},\ }\bibfield  {title} {\enquote
  {\bibinfo {title} {Generic optical excitations of correlated systems:
  $\ensuremath{\pi}$-tons},}\ }\href {\doibase 10.1103/PhysRevLett.124.047401}
  {\bibfield  {journal} {\bibinfo  {journal} {Phys. Rev. Lett.}\ }\textbf
  {\bibinfo {volume} {124}},\ \bibinfo {pages} {047401} (\bibinfo {year}
  {2020})}\BibitemShut {NoStop}%
\bibitem [{\citenamefont {Simard}\ \emph
  {et~al.}(2021{\natexlab{b}})\citenamefont {Simard}, \citenamefont
  {Takayoshi},\ and\ \citenamefont {Werner}}]{Simard}%
  \BibitemOpen
  \bibfield  {author} {\bibinfo {author} {\bibfnamefont {Olivier}\ \bibnamefont
  {Simard}}, \bibinfo {author} {\bibfnamefont {Shintaro}\ \bibnamefont
  {Takayoshi}}, \ and\ \bibinfo {author} {\bibfnamefont {Philipp}\ \bibnamefont
  {Werner}},\ }\bibfield  {title} {\enquote {\bibinfo {title} {Diagrammatic
  study of optical excitations in correlated systems},}\ }\href {\doibase
  10.1103/PhysRevB.103.104415} {\bibfield  {journal} {\bibinfo  {journal}
  {Phys. Rev. B}\ }\textbf {\bibinfo {volume} {103}},\ \bibinfo {pages}
  {104415} (\bibinfo {year} {2021}{\natexlab{b}})}\BibitemShut {NoStop}%
\bibitem [{\citenamefont {Tsuji}\ \emph {et~al.}(2009)\citenamefont {Tsuji},
  \citenamefont {Oka},\ and\ \citenamefont {Aoki}}]{Tsuji}%
  \BibitemOpen
  \bibfield  {author} {\bibinfo {author} {\bibfnamefont {Naoto}\ \bibnamefont
  {Tsuji}}, \bibinfo {author} {\bibfnamefont {Takashi}\ \bibnamefont {Oka}}, \
  and\ \bibinfo {author} {\bibfnamefont {Hideo}\ \bibnamefont {Aoki}},\
  }\bibfield  {title} {\enquote {\bibinfo {title} {Nonequilibrium steady state
  of photoexcited correlated electrons in the presence of dissipation},}\
  }\href {\doibase 10.1103/PhysRevLett.103.047403} {\bibfield  {journal}
  {\bibinfo  {journal} {Phys. Rev. Lett.}\ }\textbf {\bibinfo {volume} {103}},\
  \bibinfo {pages} {047403} (\bibinfo {year} {2009})}\BibitemShut {NoStop}%
\bibitem [{\citenamefont {Eckstein}\ and\ \citenamefont
  {Kollar}(2008)}]{Martin2008}%
  \BibitemOpen
  \bibfield  {author} {\bibinfo {author} {\bibfnamefont {Martin}\ \bibnamefont
  {Eckstein}}\ and\ \bibinfo {author} {\bibfnamefont {Marcus}\ \bibnamefont
  {Kollar}},\ }\bibfield  {title} {\enquote {\bibinfo {title} {Theory of
  time-resolved optical spectroscopy on correlated electron systems},}\ }\href
  {\doibase 10.1103/PhysRevB.78.205119} {\bibfield  {journal} {\bibinfo
  {journal} {Phys. Rev. B}\ }\textbf {\bibinfo {volume} {78}},\ \bibinfo
  {pages} {205119} (\bibinfo {year} {2008})}\BibitemShut {NoStop}%
\bibitem [{\citenamefont {Shao}\ \emph {et~al.}(2016)\citenamefont {Shao},
  \citenamefont {Tohyama}, \citenamefont {Luo},\ and\ \citenamefont
  {Lu}}]{Shao}%
  \BibitemOpen
  \bibfield  {author} {\bibinfo {author} {\bibfnamefont {Can}\ \bibnamefont
  {Shao}}, \bibinfo {author} {\bibfnamefont {Takami}\ \bibnamefont {Tohyama}},
  \bibinfo {author} {\bibfnamefont {Hong-Gang}\ \bibnamefont {Luo}}, \ and\
  \bibinfo {author} {\bibfnamefont {Hantao}\ \bibnamefont {Lu}},\ }\bibfield
  {title} {\enquote {\bibinfo {title} {Numerical method to compute optical
  conductivity based on pump-probe simulations},}\ }\href {\doibase
  10.1103/PhysRevB.93.195144} {\bibfield  {journal} {\bibinfo  {journal} {Phys.
  Rev. B}\ }\textbf {\bibinfo {volume} {93}},\ \bibinfo {pages} {195144}
  (\bibinfo {year} {2016})}\BibitemShut {NoStop}%
\bibitem [{\citenamefont {Dasari}\ \emph {et~al.}(2020)\citenamefont {Dasari},
  \citenamefont {Li}, \citenamefont {Werner},\ and\ \citenamefont
  {Eckstein}}]{Dasari}%
  \BibitemOpen
  \bibfield  {author} {\bibinfo {author} {\bibfnamefont {Nagamalleswararao}\
  \bibnamefont {Dasari}}, \bibinfo {author} {\bibfnamefont {Jiajun}\
  \bibnamefont {Li}}, \bibinfo {author} {\bibfnamefont {Philipp}\ \bibnamefont
  {Werner}}, \ and\ \bibinfo {author} {\bibfnamefont {Martin}\ \bibnamefont
  {Eckstein}},\ }\bibfield  {title} {\enquote {\bibinfo {title} {Revealing
  hund's multiplets in mott insulators under strong electric fields},}\ }\href
  {\doibase 10.1103/PhysRevB.101.161107} {\bibfield  {journal} {\bibinfo
  {journal} {Phys. Rev. B}\ }\textbf {\bibinfo {volume} {101}},\ \bibinfo
  {pages} {161107} (\bibinfo {year} {2020})}\BibitemShut {NoStop}%
\bibitem [{\citenamefont {Gole\ifmmode~\check{z}\else \v{z}\fi{}}\ \emph
  {et~al.}(2019{\natexlab{a}})\citenamefont {Gole\ifmmode~\check{z}\else
  \v{z}\fi{}}, \citenamefont {Eckstein},\ and\ \citenamefont
  {Werner}}]{PhysRevB.100.235117}%
  \BibitemOpen
  \bibfield  {author} {\bibinfo {author} {\bibfnamefont {Denis}\ \bibnamefont
  {Gole\ifmmode~\check{z}\else \v{z}\fi{}}}, \bibinfo {author} {\bibfnamefont
  {Martin}\ \bibnamefont {Eckstein}}, \ and\ \bibinfo {author} {\bibfnamefont
  {Philipp}\ \bibnamefont {Werner}},\ }\bibfield  {title} {\enquote {\bibinfo
  {title} {{Multiband nonequilibrium $GW+\text{EDMFT}$ formalism for correlated
  insulators}},}\ }\href {\doibase 10.1103/PhysRevB.100.235117} {\bibfield
  {journal} {\bibinfo  {journal} {Phys. Rev. B}\ }\textbf {\bibinfo {volume}
  {100}},\ \bibinfo {pages} {235117} (\bibinfo {year}
  {2019}{\natexlab{a}})}\BibitemShut {NoStop}%
\bibitem [{\citenamefont {Kova\ifmmode \check{c}\else
  \v{c}\fi{}evi\ifmmode~\acute{c}\else \'{c}\fi{}}\ \emph
  {et~al.}(2025)\citenamefont {Kova\ifmmode \check{c}\else
  \v{c}\fi{}evi\ifmmode~\acute{c}\else \'{c}\fi{}}, \citenamefont {Ferrero},\
  and\ \citenamefont {Vu\ifmmode \check{c}\else \v{c}\fi{}i\ifmmode
  \check{c}\else \v{c}\fi{}evi\ifmmode~\acute{c}\else \'{c}\fi{}}}]{Jaksa}%
  \BibitemOpen
  \bibfield  {author} {\bibinfo {author} {\bibfnamefont {Jeremija}\
  \bibnamefont {Kova\ifmmode \check{c}\else
  \v{c}\fi{}evi\ifmmode~\acute{c}\else \'{c}\fi{}}}, \bibinfo {author}
  {\bibfnamefont {Michel}\ \bibnamefont {Ferrero}}, \ and\ \bibinfo {author}
  {\bibfnamefont {Jak\ifmmode \check{s}\else~\v{s}\fi{}a}\ \bibnamefont
  {Vu\ifmmode \check{c}\else \v{c}\fi{}i\ifmmode \check{c}\else
  \v{c}\fi{}evi\ifmmode~\acute{c}\else \'{c}\fi{}}},\ }\bibfield  {title}
  {\enquote {\bibinfo {title} {{Toward Numerically Exact Computation of
  Conductivity in the Thermodynamic Limit of Interacting Lattice Models}},}\
  }\href {\doibase 10.1103/mm38-zttx} {\bibfield  {journal} {\bibinfo
  {journal} {Phys. Rev. Lett.}\ }\textbf {\bibinfo {volume} {135}},\ \bibinfo
  {pages} {016502} (\bibinfo {year} {2025})}\BibitemShut {NoStop}%
\bibitem [{\citenamefont {Stepanov}\ \emph {et~al.}(2019)\citenamefont
  {Stepanov}, \citenamefont {Harkov},\ and\ \citenamefont
  {Lichtenstein}}]{PhysRevB.100.205115}%
  \BibitemOpen
  \bibfield  {author} {\bibinfo {author} {\bibfnamefont {E.~A.}\ \bibnamefont
  {Stepanov}}, \bibinfo {author} {\bibfnamefont {V.}~\bibnamefont {Harkov}}, \
  and\ \bibinfo {author} {\bibfnamefont {A.~I.}\ \bibnamefont {Lichtenstein}},\
  }\bibfield  {title} {\enquote {\bibinfo {title} {{Consistent partial
  bosonization of the extended Hubbard model}},}\ }\href {\doibase
  10.1103/PhysRevB.100.205115} {\bibfield  {journal} {\bibinfo  {journal}
  {Phys. Rev. B}\ }\textbf {\bibinfo {volume} {100}},\ \bibinfo {pages}
  {205115} (\bibinfo {year} {2019})}\BibitemShut {NoStop}%
\bibitem [{\citenamefont {Harkov}\ \emph {et~al.}(2021)\citenamefont {Harkov},
  \citenamefont {Vandelli}, \citenamefont {Brener}, \citenamefont
  {Lichtenstein},\ and\ \citenamefont {Stepanov}}]{PhysRevB.103.245123}%
  \BibitemOpen
  \bibfield  {author} {\bibinfo {author} {\bibfnamefont {V.}~\bibnamefont
  {Harkov}}, \bibinfo {author} {\bibfnamefont {M.}~\bibnamefont {Vandelli}},
  \bibinfo {author} {\bibfnamefont {S.}~\bibnamefont {Brener}}, \bibinfo
  {author} {\bibfnamefont {A.~I.}\ \bibnamefont {Lichtenstein}}, \ and\
  \bibinfo {author} {\bibfnamefont {E.~A.}\ \bibnamefont {Stepanov}},\
  }\bibfield  {title} {\enquote {\bibinfo {title} {{Impact of partially
  bosonized collective fluctuations on electronic degrees of freedom}},}\
  }\href {\doibase 10.1103/PhysRevB.103.245123} {\bibfield  {journal} {\bibinfo
   {journal} {Phys. Rev. B}\ }\textbf {\bibinfo {volume} {103}},\ \bibinfo
  {pages} {245123} (\bibinfo {year} {2021})}\BibitemShut {NoStop}%
\bibitem [{\citenamefont {Vandelli}\ \emph {et~al.}(2022)\citenamefont
  {Vandelli}, \citenamefont {Kaufmann}, \citenamefont {El-Nabulsi},
  \citenamefont {Harkov}, \citenamefont {Lichtenstein},\ and\ \citenamefont
  {Stepanov}}]{10.21468/SciPostPhys.13.2.036}%
  \BibitemOpen
  \bibfield  {author} {\bibinfo {author} {\bibfnamefont {Matteo}\ \bibnamefont
  {Vandelli}}, \bibinfo {author} {\bibfnamefont {Josef}\ \bibnamefont
  {Kaufmann}}, \bibinfo {author} {\bibfnamefont {Mohammed}\ \bibnamefont
  {El-Nabulsi}}, \bibinfo {author} {\bibfnamefont {Viktor}\ \bibnamefont
  {Harkov}}, \bibinfo {author} {\bibfnamefont {Alexander~I.}\ \bibnamefont
  {Lichtenstein}}, \ and\ \bibinfo {author} {\bibfnamefont {Evgeny~A.}\
  \bibnamefont {Stepanov}},\ }\bibfield  {title} {\enquote {\bibinfo {title}
  {{Multi-band D-TRILEX approach to materials with strong electronic
  correlations}},}\ }\href {\doibase 10.21468/SciPostPhys.13.2.036} {\bibfield
  {journal} {\bibinfo  {journal} {SciPost Phys.}\ }\textbf {\bibinfo {volume}
  {13}},\ \bibinfo {pages} {036} (\bibinfo {year} {2022})}\BibitemShut
  {NoStop}%
\bibitem [{\citenamefont {Stepanov}\ \emph {et~al.}(2021)\citenamefont
  {Stepanov}, \citenamefont {Nomura}, \citenamefont {Lichtenstein},\ and\
  \citenamefont {Biermann}}]{PhysRevLett.127.207205}%
  \BibitemOpen
  \bibfield  {author} {\bibinfo {author} {\bibfnamefont {Evgeny~A.}\
  \bibnamefont {Stepanov}}, \bibinfo {author} {\bibfnamefont {Yusuke}\
  \bibnamefont {Nomura}}, \bibinfo {author} {\bibfnamefont {Alexander~I.}\
  \bibnamefont {Lichtenstein}}, \ and\ \bibinfo {author} {\bibfnamefont
  {Silke}\ \bibnamefont {Biermann}},\ }\bibfield  {title} {\enquote {\bibinfo
  {title} {{Orbital Isotropy of Magnetic Fluctuations in Correlated Electron
  Materials Induced by Hund's Exchange Coupling}},}\ }\href {\doibase
  10.1103/PhysRevLett.127.207205} {\bibfield  {journal} {\bibinfo  {journal}
  {Phys. Rev. Lett.}\ }\textbf {\bibinfo {volume} {127}},\ \bibinfo {pages}
  {207205} (\bibinfo {year} {2021})}\BibitemShut {NoStop}%
\bibitem [{\citenamefont {Stepanov}\ \emph {et~al.}(2022)\citenamefont
  {Stepanov}, \citenamefont {Harkov}, \citenamefont {R\"osner}, \citenamefont
  {Lichtenstein}, \citenamefont {Katsnelson},\ and\ \citenamefont
  {Rudenko}}]{stepanov2021coexisting}%
  \BibitemOpen
  \bibfield  {author} {\bibinfo {author} {\bibfnamefont {E.~A.}\ \bibnamefont
  {Stepanov}}, \bibinfo {author} {\bibfnamefont {V.}~\bibnamefont {Harkov}},
  \bibinfo {author} {\bibfnamefont {M.}~\bibnamefont {R\"osner}}, \bibinfo
  {author} {\bibfnamefont {A.~I.}\ \bibnamefont {Lichtenstein}}, \bibinfo
  {author} {\bibfnamefont {M.~I.}\ \bibnamefont {Katsnelson}}, \ and\ \bibinfo
  {author} {\bibfnamefont {A.~N.}\ \bibnamefont {Rudenko}},\ }\bibfield
  {title} {\enquote {\bibinfo {title} {{Coexisting charge density wave and
  ferromagnetic instabilities in monolayer InSe}},}\ }\href {\doibase
  10.1038/s41524-022-00798-4} {\bibfield  {journal} {\bibinfo  {journal} {npj
  Comput. Mater.}\ }\textbf {\bibinfo {volume} {8}},\ \bibinfo {pages} {118}
  (\bibinfo {year} {2022})}\BibitemShut {NoStop}%
\bibitem [{\citenamefont {Stepanov}(2022)}]{PhysRevLett.129.096404}%
  \BibitemOpen
  \bibfield  {author} {\bibinfo {author} {\bibfnamefont {Evgeny~A.}\
  \bibnamefont {Stepanov}},\ }\bibfield  {title} {\enquote {\bibinfo {title}
  {{Eliminating Orbital Selectivity from the Metal-Insulator Transition by
  Strong Magnetic Fluctuations}},}\ }\href {\doibase
  10.1103/PhysRevLett.129.096404} {\bibfield  {journal} {\bibinfo  {journal}
  {Phys. Rev. Lett.}\ }\textbf {\bibinfo {volume} {129}},\ \bibinfo {pages}
  {096404} (\bibinfo {year} {2022})}\BibitemShut {NoStop}%
\bibitem [{\citenamefont {Vandelli}\ \emph {et~al.}(2023)\citenamefont
  {Vandelli}, \citenamefont {Kaufmann}, \citenamefont {Harkov}, \citenamefont
  {Lichtenstein}, \citenamefont {Held},\ and\ \citenamefont
  {Stepanov}}]{PhysRevResearch.5.L022016}%
  \BibitemOpen
  \bibfield  {author} {\bibinfo {author} {\bibfnamefont {M.}~\bibnamefont
  {Vandelli}}, \bibinfo {author} {\bibfnamefont {J.}~\bibnamefont {Kaufmann}},
  \bibinfo {author} {\bibfnamefont {V.}~\bibnamefont {Harkov}}, \bibinfo
  {author} {\bibfnamefont {A.~I.}\ \bibnamefont {Lichtenstein}}, \bibinfo
  {author} {\bibfnamefont {K.}~\bibnamefont {Held}}, \ and\ \bibinfo {author}
  {\bibfnamefont {E.~A.}\ \bibnamefont {Stepanov}},\ }\bibfield  {title}
  {\enquote {\bibinfo {title} {{Extended regime of metastable metallic and
  insulating phases in a two-orbital electronic system}},}\ }\href {\doibase
  10.1103/PhysRevResearch.5.L022016} {\bibfield  {journal} {\bibinfo  {journal}
  {Phys. Rev. Res.}\ }\textbf {\bibinfo {volume} {5}},\ \bibinfo {pages}
  {L022016} (\bibinfo {year} {2023})}\BibitemShut {NoStop}%
\bibitem [{\citenamefont {Vandelli}\ \emph {et~al.}(2024)\citenamefont
  {Vandelli}, \citenamefont {Galler}, \citenamefont {Rubio}, \citenamefont
  {Lichtenstein}, \citenamefont {Biermann},\ and\ \citenamefont
  {Stepanov}}]{vandelli2024doping}%
  \BibitemOpen
  \bibfield  {author} {\bibinfo {author} {\bibfnamefont {M.}~\bibnamefont
  {Vandelli}}, \bibinfo {author} {\bibfnamefont {A.}~\bibnamefont {Galler}},
  \bibinfo {author} {\bibfnamefont {A.}~\bibnamefont {Rubio}}, \bibinfo
  {author} {\bibfnamefont {A.~I.}\ \bibnamefont {Lichtenstein}}, \bibinfo
  {author} {\bibfnamefont {S.}~\bibnamefont {Biermann}}, \ and\ \bibinfo
  {author} {\bibfnamefont {E.~A.}\ \bibnamefont {Stepanov}},\ }\bibfield
  {title} {\enquote {\bibinfo {title} {{Doping-dependent charge- and
  spin-density wave orderings in a monolayer of Pb adatoms on Si(111)}},}\
  }\href {\doibase 10.1038/s41535-024-00630-w} {\bibfield  {journal} {\bibinfo
  {journal} {npj Quantum Mater.}\ }\textbf {\bibinfo {volume} {9}},\ \bibinfo
  {pages} {19} (\bibinfo {year} {2024})}\BibitemShut {NoStop}%
\bibitem [{\citenamefont {Stepanov}\ and\ \citenamefont
  {Biermann}(2024)}]{PhysRevLett.132.226501}%
  \BibitemOpen
  \bibfield  {author} {\bibinfo {author} {\bibfnamefont {Evgeny~A.}\
  \bibnamefont {Stepanov}}\ and\ \bibinfo {author} {\bibfnamefont {Silke}\
  \bibnamefont {Biermann}},\ }\bibfield  {title} {\enquote {\bibinfo {title}
  {{Can Orbital-Selective N\'eel Transitions Survive Strong Nonlocal Electronic
  Correlations?}}}\ }\href {\doibase 10.1103/PhysRevLett.132.226501} {\bibfield
   {journal} {\bibinfo  {journal} {Phys. Rev. Lett.}\ }\textbf {\bibinfo
  {volume} {132}},\ \bibinfo {pages} {226501} (\bibinfo {year}
  {2024})}\BibitemShut {NoStop}%
\bibitem [{\citenamefont {Stepanov}\ \emph
  {et~al.}(2024{\natexlab{a}})\citenamefont {Stepanov}, \citenamefont
  {Vandelli}, \citenamefont {Lichtenstein},\ and\ \citenamefont
  {Lechermann}}]{stepanov2023charge}%
  \BibitemOpen
  \bibfield  {author} {\bibinfo {author} {\bibfnamefont {Evgeny~A.}\
  \bibnamefont {Stepanov}}, \bibinfo {author} {\bibfnamefont {Matteo}\
  \bibnamefont {Vandelli}}, \bibinfo {author} {\bibfnamefont {Alexander~I.}\
  \bibnamefont {Lichtenstein}}, \ and\ \bibinfo {author} {\bibfnamefont
  {Frank}\ \bibnamefont {Lechermann}},\ }\bibfield  {title} {\enquote {\bibinfo
  {title} {{Charge Density Wave Ordering in NdNiO$_2$: Effects of Multiorbital
  Nonlocal Correlations}},}\ }\href {\doibase 10.1038/s41524-024-01298-3}
  {\bibfield  {journal} {\bibinfo  {journal} {npj Comput. Mater.}\ }\textbf
  {\bibinfo {volume} {10}},\ \bibinfo {pages} {108} (\bibinfo {year}
  {2024}{\natexlab{a}})}\BibitemShut {NoStop}%
\bibitem [{\citenamefont {Stepanov}\ \emph
  {et~al.}(2024{\natexlab{b}})\citenamefont {Stepanov}, \citenamefont
  {Chatzieleftheriou}, \citenamefont {Wagner},\ and\ \citenamefont
  {Sangiovanni}}]{PhysRevB.110.L161106}%
  \BibitemOpen
  \bibfield  {author} {\bibinfo {author} {\bibfnamefont {Evgeny~A.}\
  \bibnamefont {Stepanov}}, \bibinfo {author} {\bibfnamefont {Maria}\
  \bibnamefont {Chatzieleftheriou}}, \bibinfo {author} {\bibfnamefont {Niklas}\
  \bibnamefont {Wagner}}, \ and\ \bibinfo {author} {\bibfnamefont {Giorgio}\
  \bibnamefont {Sangiovanni}},\ }\bibfield  {title} {\enquote {\bibinfo {title}
  {{Interconnected renormalization of Hubbard bands and Green's function zeros
  in Mott insulators induced by strong magnetic fluctuations}},}\ }\href
  {\doibase 10.1103/PhysRevB.110.L161106} {\bibfield  {journal} {\bibinfo
  {journal} {Phys. Rev. B}\ }\textbf {\bibinfo {volume} {110}},\ \bibinfo
  {pages} {L161106} (\bibinfo {year} {2024}{\natexlab{b}})}\BibitemShut
  {NoStop}%
\bibitem [{\citenamefont {Stepanov}(2025)}]{j6bj-gz7j}%
  \BibitemOpen
  \bibfield  {author} {\bibinfo {author} {\bibfnamefont {Evgeny~A.}\
  \bibnamefont {Stepanov}},\ }\bibfield  {title} {\enquote {\bibinfo {title}
  {{Fingerprints of a charge ice state in the doped Mott insulator
  ${\mathrm{Nb}}_{3}{\mathrm{Cl}}_{8}$}},}\ }\href {\doibase 10.1103/j6bj-gz7j}
  {\bibfield  {journal} {\bibinfo  {journal} {Phys. Rev. B}\ }\textbf {\bibinfo
  {volume} {112}},\ \bibinfo {pages} {045131} (\bibinfo {year}
  {2025})}\BibitemShut {NoStop}%
\bibitem [{\citenamefont {Chatzieleftheriou}\ \emph {et~al.}(2025)\citenamefont
  {Chatzieleftheriou}, \citenamefont {Rudenko}, \citenamefont {Sidis},
  \citenamefont {Biermann},\ and\ \citenamefont {Stepanov}}]{Ruthenates}%
  \BibitemOpen
  \bibfield  {author} {\bibinfo {author} {\bibfnamefont {Maria}\ \bibnamefont
  {Chatzieleftheriou}}, \bibinfo {author} {\bibfnamefont {Alexander~N.}\
  \bibnamefont {Rudenko}}, \bibinfo {author} {\bibfnamefont {Yvan}\
  \bibnamefont {Sidis}}, \bibinfo {author} {\bibfnamefont {Silke}\ \bibnamefont
  {Biermann}}, \ and\ \bibinfo {author} {\bibfnamefont {Evgeny~A.}\
  \bibnamefont {Stepanov}},\ }\bibfield  {title} {\enquote {\bibinfo {title}
  {{Nature of momentum- and orbital-dependent magnetic fluctuations in
  ${\mathrm{Sr}}_{2}{\mathrm{RuO}}_{4}$}},}\ }\href {\doibase
  10.1103/ts6y-zb6m} {\bibfield  {journal} {\bibinfo  {journal} {Phys. Rev. B}\
  }\textbf {\bibinfo {volume} {112}},\ \bibinfo {pages} {195118} (\bibinfo
  {year} {2025})}\BibitemShut {NoStop}%
\bibitem [{\citenamefont {Stepanov}\ \emph {et~al.}(2026)\citenamefont
  {Stepanov}, \citenamefont {Iskakov}, \citenamefont {Katsnelson},\ and\
  \citenamefont {Lichtenstein}}]{Cuprates}%
  \BibitemOpen
  \bibfield  {author} {\bibinfo {author} {\bibfnamefont {E.~A.}\ \bibnamefont
  {Stepanov}}, \bibinfo {author} {\bibfnamefont {S.}~\bibnamefont {Iskakov}},
  \bibinfo {author} {\bibfnamefont {M.~I.}\ \bibnamefont {Katsnelson}}, \ and\
  \bibinfo {author} {\bibfnamefont {A.~I.}\ \bibnamefont {Lichtenstein}},\
  }\bibfield  {title} {\enquote {\bibinfo {title} {{Superconductivity of bad
  fermions and the origin of two gaps in cuprates}},}\ }\href {\doibase
  10.1038/s42005-026-02532-8} {\bibfield  {journal} {\bibinfo  {journal}
  {Commun. Phys.}\ } (\bibinfo {year} {2026}),\
  10.1038/s42005-026-02532-8}\BibitemShut {NoStop}%
\bibitem [{\citenamefont {Gole\ifmmode~\check{z}\else \v{z}\fi{}}\ \emph
  {et~al.}(2017)\citenamefont {Gole\ifmmode~\check{z}\else \v{z}\fi{}},
  \citenamefont {Boehnke}, \citenamefont {Strand}, \citenamefont {Eckstein},\
  and\ \citenamefont {Werner}}]{PhysRevLett.118.246402}%
  \BibitemOpen
  \bibfield  {author} {\bibinfo {author} {\bibfnamefont {Denis}\ \bibnamefont
  {Gole\ifmmode~\check{z}\else \v{z}\fi{}}}, \bibinfo {author} {\bibfnamefont
  {Lewin}\ \bibnamefont {Boehnke}}, \bibinfo {author} {\bibfnamefont {Hugo
  U.~R.}\ \bibnamefont {Strand}}, \bibinfo {author} {\bibfnamefont {Martin}\
  \bibnamefont {Eckstein}}, \ and\ \bibinfo {author} {\bibfnamefont {Philipp}\
  \bibnamefont {Werner}},\ }\bibfield  {title} {\enquote {\bibinfo {title}
  {{Nonequilibrium $GW+\mathrm{EDMFT}$: Antiscreening and Inverted Populations
  from Nonlocal Correlations}},}\ }\href {\doibase
  10.1103/PhysRevLett.118.246402} {\bibfield  {journal} {\bibinfo  {journal}
  {Phys. Rev. Lett.}\ }\textbf {\bibinfo {volume} {118}},\ \bibinfo {pages}
  {246402} (\bibinfo {year} {2017})}\BibitemShut {NoStop}%
\bibitem [{\citenamefont {Gole\ifmmode~\check{z}\else \v{z}\fi{}}\ \emph
  {et~al.}(2019{\natexlab{b}})\citenamefont {Gole\ifmmode~\check{z}\else
  \v{z}\fi{}}, \citenamefont {Boehnke}, \citenamefont {Eckstein},\ and\
  \citenamefont {Werner}}]{PhysRevB.100.041111}%
  \BibitemOpen
  \bibfield  {author} {\bibinfo {author} {\bibfnamefont {Denis}\ \bibnamefont
  {Gole\ifmmode~\check{z}\else \v{z}\fi{}}}, \bibinfo {author} {\bibfnamefont
  {Lewin}\ \bibnamefont {Boehnke}}, \bibinfo {author} {\bibfnamefont {Martin}\
  \bibnamefont {Eckstein}}, \ and\ \bibinfo {author} {\bibfnamefont {Philipp}\
  \bibnamefont {Werner}},\ }\bibfield  {title} {\enquote {\bibinfo {title}
  {{Dynamics of photodoped charge transfer insulators}},}\ }\href {\doibase
  10.1103/PhysRevB.100.041111} {\bibfield  {journal} {\bibinfo  {journal}
  {Phys. Rev. B}\ }\textbf {\bibinfo {volume} {100}},\ \bibinfo {pages}
  {041111} (\bibinfo {year} {2019}{\natexlab{b}})}\BibitemShut {NoStop}%
\bibitem [{\citenamefont {Geng}\ \emph {et~al.}(2025)\citenamefont {Geng},
  \citenamefont {Yan},\ and\ \citenamefont {Werner}}]{Geng}%
  \BibitemOpen
  \bibfield  {author} {\bibinfo {author} {\bibfnamefont {Lei}\ \bibnamefont
  {Geng}}, \bibinfo {author} {\bibfnamefont {Jiawei}\ \bibnamefont {Yan}}, \
  and\ \bibinfo {author} {\bibfnamefont {Philipp}\ \bibnamefont {Werner}},\
  }\bibfield  {title} {\enquote {\bibinfo {title} {{Two-particle
  self-consistent approach combined with dynamical mean field theory: A
  real-frequency study of the square-lattice Hubbard model}},}\ }\href
  {\doibase 10.1103/PhysRevB.111.115143} {\bibfield  {journal} {\bibinfo
  {journal} {Phys. Rev. B}\ }\textbf {\bibinfo {volume} {111}},\ \bibinfo
  {pages} {115143} (\bibinfo {year} {2025})}\BibitemShut {NoStop}%
\bibitem [{\citenamefont {Charlebois}\ and\ \citenamefont
  {Imada}(2020)}]{PhysRevX.10.041023}%
  \BibitemOpen
  \bibfield  {author} {\bibinfo {author} {\bibfnamefont {Maxime}\ \bibnamefont
  {Charlebois}}\ and\ \bibinfo {author} {\bibfnamefont {Masatoshi}\
  \bibnamefont {Imada}},\ }\bibfield  {title} {\enquote {\bibinfo {title}
  {{Single-Particle Spectral Function Formulated and Calculated by Variational
  Monte Carlo Method with Application to $d$-Wave Superconducting State}},}\
  }\href {\doibase 10.1103/PhysRevX.10.041023} {\bibfield  {journal} {\bibinfo
  {journal} {Phys. Rev. X}\ }\textbf {\bibinfo {volume} {10}},\ \bibinfo
  {pages} {041023} (\bibinfo {year} {2020})}\BibitemShut {NoStop}%
\bibitem [{\citenamefont {Singh}\ \emph {et~al.}(2022)\citenamefont {Singh},
  \citenamefont {Huang}, \citenamefont {Xie}, \citenamefont {Okamoto},
  \citenamefont {Chen}, \citenamefont {Watanabe}, \citenamefont {Fujimori},
  \citenamefont {Imada},\ and\ \citenamefont
  {Huang}}]{singh2022unconventional}%
  \BibitemOpen
  \bibfield  {author} {\bibinfo {author} {\bibfnamefont {A.}~\bibnamefont
  {Singh}}, \bibinfo {author} {\bibfnamefont {H.~Y.}\ \bibnamefont {Huang}},
  \bibinfo {author} {\bibfnamefont {J.~D.}\ \bibnamefont {Xie}}, \bibinfo
  {author} {\bibfnamefont {J.}~\bibnamefont {Okamoto}}, \bibinfo {author}
  {\bibfnamefont {C.~T.}\ \bibnamefont {Chen}}, \bibinfo {author}
  {\bibfnamefont {T.}~\bibnamefont {Watanabe}}, \bibinfo {author}
  {\bibfnamefont {A.}~\bibnamefont {Fujimori}}, \bibinfo {author}
  {\bibfnamefont {M.}~\bibnamefont {Imada}}, \ and\ \bibinfo {author}
  {\bibfnamefont {D.~J.}\ \bibnamefont {Huang}},\ }\bibfield  {title} {\enquote
  {\bibinfo {title} {{Unconventional exciton evolution from the pseudogap to
  superconducting phases in cuprates}},}\ }\href {\doibase
  10.1038/s41467-022-35210-8} {\bibfield  {journal} {\bibinfo  {journal} {Nat.
  Commun.}\ }\textbf {\bibinfo {volume} {13}},\ \bibinfo {pages} {7906}
  (\bibinfo {year} {2022})}\BibitemShut {NoStop}%
\bibitem [{\citenamefont {Kim}\ \emph {et~al.}(2008)\citenamefont {Kim},
  \citenamefont {Jin}, \citenamefont {Moon}, \citenamefont {Kim}, \citenamefont
  {Park}, \citenamefont {Leem}, \citenamefont {Yu}, \citenamefont {Noh},
  \citenamefont {Kim}, \citenamefont {Oh}, \citenamefont {Park}, \citenamefont
  {Durairaj}, \citenamefont {Cao},\ and\ \citenamefont
  {Rotenberg}}]{PhysRevLett.101.076402}%
  \BibitemOpen
  \bibfield  {author} {\bibinfo {author} {\bibfnamefont {B.~J.}\ \bibnamefont
  {Kim}}, \bibinfo {author} {\bibfnamefont {Hosub}\ \bibnamefont {Jin}},
  \bibinfo {author} {\bibfnamefont {S.~J.}\ \bibnamefont {Moon}}, \bibinfo
  {author} {\bibfnamefont {J.-Y.}\ \bibnamefont {Kim}}, \bibinfo {author}
  {\bibfnamefont {B.-G.}\ \bibnamefont {Park}}, \bibinfo {author}
  {\bibfnamefont {C.~S.}\ \bibnamefont {Leem}}, \bibinfo {author}
  {\bibfnamefont {Jaejun}\ \bibnamefont {Yu}}, \bibinfo {author} {\bibfnamefont
  {T.~W.}\ \bibnamefont {Noh}}, \bibinfo {author} {\bibfnamefont
  {C.}~\bibnamefont {Kim}}, \bibinfo {author} {\bibfnamefont {S.-J.}\
  \bibnamefont {Oh}}, \bibinfo {author} {\bibfnamefont {J.-H.}\ \bibnamefont
  {Park}}, \bibinfo {author} {\bibfnamefont {V.}~\bibnamefont {Durairaj}},
  \bibinfo {author} {\bibfnamefont {G.}~\bibnamefont {Cao}}, \ and\ \bibinfo
  {author} {\bibfnamefont {E.}~\bibnamefont {Rotenberg}},\ }\bibfield  {title}
  {\enquote {\bibinfo {title} {{Novel ${J}_{\mathrm{eff}}=1/2$ Mott State
  Induced by Relativistic Spin-Orbit Coupling in
  ${\mathrm{Sr}}_{2}{\mathrm{IrO}}_{4}$}},}\ }\href {\doibase
  10.1103/PhysRevLett.101.076402} {\bibfield  {journal} {\bibinfo  {journal}
  {Phys. Rev. Lett.}\ }\textbf {\bibinfo {volume} {101}},\ \bibinfo {pages}
  {076402} (\bibinfo {year} {2008})}\BibitemShut {NoStop}%
\bibitem [{\citenamefont {Moon}\ \emph {et~al.}(2009)\citenamefont {Moon},
  \citenamefont {Jin}, \citenamefont {Choi}, \citenamefont {Lee}, \citenamefont
  {Seo}, \citenamefont {Yu}, \citenamefont {Cao}, \citenamefont {Noh},\ and\
  \citenamefont {Lee}}]{PhysRevB.80.195110}%
  \BibitemOpen
  \bibfield  {author} {\bibinfo {author} {\bibfnamefont {S.~J.}\ \bibnamefont
  {Moon}}, \bibinfo {author} {\bibfnamefont {Hosub}\ \bibnamefont {Jin}},
  \bibinfo {author} {\bibfnamefont {W.~S.}\ \bibnamefont {Choi}}, \bibinfo
  {author} {\bibfnamefont {J.~S.}\ \bibnamefont {Lee}}, \bibinfo {author}
  {\bibfnamefont {S.~S.~A.}\ \bibnamefont {Seo}}, \bibinfo {author}
  {\bibfnamefont {J.}~\bibnamefont {Yu}}, \bibinfo {author} {\bibfnamefont
  {G.}~\bibnamefont {Cao}}, \bibinfo {author} {\bibfnamefont {T.~W.}\
  \bibnamefont {Noh}}, \ and\ \bibinfo {author} {\bibfnamefont {Y.~S.}\
  \bibnamefont {Lee}},\ }\bibfield  {title} {\enquote {\bibinfo {title}
  {{Temperature dependence of the electronic structure of the
  ${J}_{\text{eff}}=\frac{1}{2}$ Mott insulator
  ${\text{Sr}}_{2}{\text{IrO}}_{4}$ studied by optical spectroscopy}},}\ }\href
  {\doibase 10.1103/PhysRevB.80.195110} {\bibfield  {journal} {\bibinfo
  {journal} {Phys. Rev. B}\ }\textbf {\bibinfo {volume} {80}},\ \bibinfo
  {pages} {195110} (\bibinfo {year} {2009})}\BibitemShut {NoStop}%
\bibitem [{\citenamefont {Seo}\ \emph {et~al.}(2017)\citenamefont {Seo},
  \citenamefont {Ahn}, \citenamefont {Song}, \citenamefont {Chen},
  \citenamefont {Wilson},\ and\ \citenamefont {Moon}}]{seo2017infrared}%
  \BibitemOpen
  \bibfield  {author} {\bibinfo {author} {\bibfnamefont {J.~H.}\ \bibnamefont
  {Seo}}, \bibinfo {author} {\bibfnamefont {G.~H.}\ \bibnamefont {Ahn}},
  \bibinfo {author} {\bibfnamefont {S.~J.}\ \bibnamefont {Song}}, \bibinfo
  {author} {\bibfnamefont {X.}~\bibnamefont {Chen}}, \bibinfo {author}
  {\bibfnamefont {S.~D.}\ \bibnamefont {Wilson}}, \ and\ \bibinfo {author}
  {\bibfnamefont {S.~J.}\ \bibnamefont {Moon}},\ }\bibfield  {title} {\enquote
  {\bibinfo {title} {{Infrared probe of pseudogap in electron-doped
  Sr$_2$IrO$_4$}},}\ }\href {\doibase 10.1038/s41598-017-10725-z} {\bibfield
  {journal} {\bibinfo  {journal} {Sci. Rep.}\ }\textbf {\bibinfo {volume}
  {7}},\ \bibinfo {pages} {10494} (\bibinfo {year} {2017})}\BibitemShut
  {NoStop}%
\bibitem [{\citenamefont {Nichols}\ \emph {et~al.}(2014)\citenamefont
  {Nichols}, \citenamefont {Korneta}, \citenamefont {Terzic}, \citenamefont
  {Cao}, \citenamefont {Brill},\ and\ \citenamefont {Seo}}]{10.1063/1.4870049}%
  \BibitemOpen
  \bibfield  {author} {\bibinfo {author} {\bibfnamefont {J.}~\bibnamefont
  {Nichols}}, \bibinfo {author} {\bibfnamefont {O.~B.}\ \bibnamefont
  {Korneta}}, \bibinfo {author} {\bibfnamefont {J.}~\bibnamefont {Terzic}},
  \bibinfo {author} {\bibfnamefont {G.}~\bibnamefont {Cao}}, \bibinfo {author}
  {\bibfnamefont {J.~W.}\ \bibnamefont {Brill}}, \ and\ \bibinfo {author}
  {\bibfnamefont {S.~S.~A.}\ \bibnamefont {Seo}},\ }\bibfield  {title}
  {\enquote {\bibinfo {title} {{Epitaxial Ba$_2$IrO$_4$ thin-films grown on
  SrTiO3 substrates by pulsed laser deposition}},}\ }\href {\doibase
  10.1063/1.4870049} {\bibfield  {journal} {\bibinfo  {journal} {Appl. Phys.
  Lett.}\ }\textbf {\bibinfo {volume} {104}},\ \bibinfo {pages} {121913}
  (\bibinfo {year} {2014})}\BibitemShut {NoStop}%
\bibitem [{\citenamefont {Cassol}\ \emph {et~al.}(2025)\citenamefont {Cassol},
  \citenamefont {Gaspard}, \citenamefont {Martins}, \citenamefont {Casula},\
  and\ \citenamefont {Lenz}}]{Cassol}%
  \BibitemOpen
  \bibfield  {author} {\bibinfo {author} {\bibfnamefont {Francesco}\
  \bibnamefont {Cassol}}, \bibinfo {author} {\bibfnamefont {Léo}\ \bibnamefont
  {Gaspard}}, \bibinfo {author} {\bibfnamefont {Cyril}\ \bibnamefont
  {Martins}}, \bibinfo {author} {\bibfnamefont {Michele}\ \bibnamefont
  {Casula}}, \ and\ \bibinfo {author} {\bibfnamefont {Benjamin}\ \bibnamefont
  {Lenz}},\ }\href@noop {} {\enquote {\bibinfo {title} {Spin-polaron
  fingerprints in the optical conductivity of iridates},}\ }\bibinfo
  {howpublished} {To be published} (\bibinfo {year} {2025})\BibitemShut
  {NoStop}%
\bibitem [{\citenamefont {Lenz}\ \emph {et~al.}(2019)\citenamefont {Lenz},
  \citenamefont {Martins},\ and\ \citenamefont {Biermann}}]{Lenz_2019}%
  \BibitemOpen
  \bibfield  {author} {\bibinfo {author} {\bibfnamefont {B.}~\bibnamefont
  {Lenz}}, \bibinfo {author} {\bibfnamefont {C.}~\bibnamefont {Martins}}, \
  and\ \bibinfo {author} {\bibfnamefont {S.}~\bibnamefont {Biermann}},\
  }\bibfield  {title} {\enquote {\bibinfo {title} {{Spectral functions of
  Sr$_2$IrO$_4$: theory versus experiment}},}\ }\href {\doibase
  10.1088/1361-648X/ab146a} {\bibfield  {journal} {\bibinfo  {journal} {J.
  Phys.: Condens. Matter}\ }\textbf {\bibinfo {volume} {31}},\ \bibinfo {pages}
  {293001} (\bibinfo {year} {2019})}\BibitemShut {NoStop}%
\bibitem [{\citenamefont {Keimer}\ \emph {et~al.}(2015)\citenamefont {Keimer},
  \citenamefont {Kivelson}, \citenamefont {Norman}, \citenamefont {Uchida},\
  and\ \citenamefont {Zaanen}}]{keimer2015quantum}%
  \BibitemOpen
  \bibfield  {author} {\bibinfo {author} {\bibfnamefont {Bernhard}\
  \bibnamefont {Keimer}}, \bibinfo {author} {\bibfnamefont {Steven~A.}\
  \bibnamefont {Kivelson}}, \bibinfo {author} {\bibfnamefont {Michael~R.}\
  \bibnamefont {Norman}}, \bibinfo {author} {\bibfnamefont {Shinichi}\
  \bibnamefont {Uchida}}, \ and\ \bibinfo {author} {\bibfnamefont
  {J.}~\bibnamefont {Zaanen}},\ }\bibfield  {title} {\enquote {\bibinfo {title}
  {{From quantum matter to high-temperature superconductivity in copper
  oxides}},}\ }\href {\doibase 10.1038/nature14165} {\bibfield  {journal}
  {\bibinfo  {journal} {Nature}\ }\textbf {\bibinfo {volume} {518}},\ \bibinfo
  {pages} {179--186} (\bibinfo {year} {2015})}\BibitemShut {NoStop}%
\bibitem [{\citenamefont {Phillips}\ \emph {et~al.}(2022)\citenamefont
  {Phillips}, \citenamefont {Hussey},\ and\ \citenamefont
  {Abbamonte}}]{doi:10.1126/science.abh4273}%
  \BibitemOpen
  \bibfield  {author} {\bibinfo {author} {\bibfnamefont {Philip~W.}\
  \bibnamefont {Phillips}}, \bibinfo {author} {\bibfnamefont {Nigel~E.}\
  \bibnamefont {Hussey}}, \ and\ \bibinfo {author} {\bibfnamefont {Peter}\
  \bibnamefont {Abbamonte}},\ }\bibfield  {title} {\enquote {\bibinfo {title}
  {Stranger than metals},}\ }\href {\doibase 10.1126/science.abh4273}
  {\bibfield  {journal} {\bibinfo  {journal} {Science}\ }\textbf {\bibinfo
  {volume} {377}},\ \bibinfo {pages} {eabh4273} (\bibinfo {year}
  {2022})}\BibitemShut {NoStop}%
\bibitem [{\citenamefont {Eckstein}\ and\ \citenamefont
  {Werner}(2010)}]{PhysRevB.82.115115}%
  \BibitemOpen
  \bibfield  {author} {\bibinfo {author} {\bibfnamefont {Martin}\ \bibnamefont
  {Eckstein}}\ and\ \bibinfo {author} {\bibfnamefont {Philipp}\ \bibnamefont
  {Werner}},\ }\bibfield  {title} {\enquote {\bibinfo {title} {Nonequilibrium
  dynamical mean-field calculations based on the noncrossing approximation and
  its generalizations},}\ }\href {\doibase 10.1103/PhysRevB.82.115115}
  {\bibfield  {journal} {\bibinfo  {journal} {Phys. Rev. B}\ }\textbf {\bibinfo
  {volume} {82}},\ \bibinfo {pages} {115115} (\bibinfo {year}
  {2010})}\BibitemShut {NoStop}%
\bibitem [{\citenamefont {Schüler}\ \emph {et~al.}(2020)\citenamefont
  {Schüler}, \citenamefont {Golež}, \citenamefont {Murakami}, \citenamefont
  {Bittner}, \citenamefont {Herrmann}, \citenamefont {Strand}, \citenamefont
  {Werner},\ and\ \citenamefont {Eckstein}}]{SCHULER2020107484}%
  \BibitemOpen
  \bibfield  {author} {\bibinfo {author} {\bibfnamefont {Michael}\ \bibnamefont
  {Schüler}}, \bibinfo {author} {\bibfnamefont {Denis}\ \bibnamefont
  {Golež}}, \bibinfo {author} {\bibfnamefont {Yuta}\ \bibnamefont {Murakami}},
  \bibinfo {author} {\bibfnamefont {Nikolaj}\ \bibnamefont {Bittner}}, \bibinfo
  {author} {\bibfnamefont {Andreas}\ \bibnamefont {Herrmann}}, \bibinfo
  {author} {\bibfnamefont {Hugo~U.R.}\ \bibnamefont {Strand}}, \bibinfo
  {author} {\bibfnamefont {Philipp}\ \bibnamefont {Werner}}, \ and\ \bibinfo
  {author} {\bibfnamefont {Martin}\ \bibnamefont {Eckstein}},\ }\bibfield
  {title} {\enquote {\bibinfo {title} {{NESSi: The Non-Equilibrium Systems
  Simulation package}},}\ }\href {\doibase
  https://doi.org/10.1016/j.cpc.2020.107484} {\bibfield  {journal} {\bibinfo
  {journal} {Comput. Phys. Commun.}\ }\textbf {\bibinfo {volume} {257}},\
  \bibinfo {pages} {107484} (\bibinfo {year} {2020})}\BibitemShut {NoStop}%
\end{thebibliography}%


\begin{thebibliography}{12}%
\makeatletter
\providecommand \@ifxundefined [1]{%
 \@ifx{#1\undefined}
}%
\providecommand \@ifnum [1]{%
 \ifnum #1\expandafter \@firstoftwo
 \else \expandafter \@secondoftwo
 \fi
}%
\providecommand \@ifx [1]{%
 \ifx #1\expandafter \@firstoftwo
 \else \expandafter \@secondoftwo
 \fi
}%
\providecommand \natexlab [1]{#1}%
\providecommand \enquote  [1]{``#1''}%
\providecommand \bibnamefont  [1]{#1}%
\providecommand \bibfnamefont [1]{#1}%
\providecommand \citenamefont [1]{#1}%
\providecommand \href@noop [0]{\@secondoftwo}%
\providecommand \href [0]{\begingroup \@sanitize@url \@href}%
\providecommand \@href[1]{\@@startlink{#1}\@@href}%
\providecommand \@@href[1]{\endgroup#1\@@endlink}%
\providecommand \@sanitize@url [0]{\catcode `\\12\catcode `\$12\catcode
  `\&12\catcode `\#12\catcode `\^12\catcode `\_12\catcode `\%12\relax}%
\providecommand \@@startlink[1]{}%
\providecommand \@@endlink[0]{}%
\providecommand \url  [0]{\begingroup\@sanitize@url \@url }%
\providecommand \@url [1]{\endgroup\@href {#1}{\urlprefix }}%
\providecommand \urlprefix  [0]{URL }%
\providecommand \Eprint [0]{\href }%
\providecommand \doibase [0]{http://dx.doi.org/}%
\providecommand \selectlanguage [0]{\@gobble}%
\providecommand \bibinfo  [0]{\@secondoftwo}%
\providecommand \bibfield  [0]{\@secondoftwo}%
\providecommand \translation [1]{[#1]}%
\providecommand \BibitemOpen [0]{}%
\providecommand \bibitemStop [0]{}%
\providecommand \bibitemNoStop [0]{.\EOS\space}%
\providecommand \EOS [0]{\spacefactor3000\relax}%
\providecommand \BibitemShut  [1]{\csname bibitem#1\endcsname}%
\let\auto@bib@innerbib\@empty
\bibitem [{\citenamefont {Stepanov}\ \emph {et~al.}(2019)\citenamefont
  {Stepanov}, \citenamefont {Harkov},\ and\ \citenamefont
  {Lichtenstein}}]{PhysRevB.100.205115}%
  \BibitemOpen
  \bibfield  {author} {\bibinfo {author} {\bibfnamefont {E.~A.}\ \bibnamefont
  {Stepanov}}, \bibinfo {author} {\bibfnamefont {V.}~\bibnamefont {Harkov}}, \
  and\ \bibinfo {author} {\bibfnamefont {A.~I.}\ \bibnamefont {Lichtenstein}},\
  }\bibfield  {title} {\enquote {\bibinfo {title} {{Consistent partial
  bosonization of the extended Hubbard model}},}\ }\href {\doibase
  10.1103/PhysRevB.100.205115} {\bibfield  {journal} {\bibinfo  {journal}
  {Phys. Rev. B}\ }\textbf {\bibinfo {volume} {100}},\ \bibinfo {pages}
  {205115} (\bibinfo {year} {2019})}\BibitemShut {NoStop}%
\bibitem [{\citenamefont {Harkov}\ \emph {et~al.}(2021)\citenamefont {Harkov},
  \citenamefont {Vandelli}, \citenamefont {Brener}, \citenamefont
  {Lichtenstein},\ and\ \citenamefont {Stepanov}}]{PhysRevB.103.245123}%
  \BibitemOpen
  \bibfield  {author} {\bibinfo {author} {\bibfnamefont {V.}~\bibnamefont
  {Harkov}}, \bibinfo {author} {\bibfnamefont {M.}~\bibnamefont {Vandelli}},
  \bibinfo {author} {\bibfnamefont {S.}~\bibnamefont {Brener}}, \bibinfo
  {author} {\bibfnamefont {A.~I.}\ \bibnamefont {Lichtenstein}}, \ and\
  \bibinfo {author} {\bibfnamefont {E.~A.}\ \bibnamefont {Stepanov}},\
  }\bibfield  {title} {\enquote {\bibinfo {title} {{Impact of partially
  bosonized collective fluctuations on electronic degrees of freedom}},}\
  }\href {\doibase 10.1103/PhysRevB.103.245123} {\bibfield  {journal} {\bibinfo
   {journal} {Phys. Rev. B}\ }\textbf {\bibinfo {volume} {103}},\ \bibinfo
  {pages} {245123} (\bibinfo {year} {2021})}\BibitemShut {NoStop}%
\bibitem [{\citenamefont {Vandelli}\ \emph {et~al.}(2022)\citenamefont
  {Vandelli}, \citenamefont {Kaufmann}, \citenamefont {El-Nabulsi},
  \citenamefont {Harkov}, \citenamefont {Lichtenstein},\ and\ \citenamefont
  {Stepanov}}]{10.21468/SciPostPhys.13.2.036}%
  \BibitemOpen
  \bibfield  {author} {\bibinfo {author} {\bibfnamefont {Matteo}\ \bibnamefont
  {Vandelli}}, \bibinfo {author} {\bibfnamefont {Josef}\ \bibnamefont
  {Kaufmann}}, \bibinfo {author} {\bibfnamefont {Mohammed}\ \bibnamefont
  {El-Nabulsi}}, \bibinfo {author} {\bibfnamefont {Viktor}\ \bibnamefont
  {Harkov}}, \bibinfo {author} {\bibfnamefont {Alexander~I.}\ \bibnamefont
  {Lichtenstein}}, \ and\ \bibinfo {author} {\bibfnamefont {Evgeny~A.}\
  \bibnamefont {Stepanov}},\ }\bibfield  {title} {\enquote {\bibinfo {title}
  {{Multi-band D-TRILEX approach to materials with strong electronic
  correlations}},}\ }\href {\doibase 10.21468/SciPostPhys.13.2.036} {\bibfield
  {journal} {\bibinfo  {journal} {SciPost Phys.}\ }\textbf {\bibinfo {volume}
  {13}},\ \bibinfo {pages} {036} (\bibinfo {year} {2022})}\BibitemShut
  {NoStop}%
\bibitem [{\citenamefont {Dasari}\ \emph {et~al.}(2025)\citenamefont {Dasari},
  \citenamefont {Strand}, \citenamefont {Eckstein}, \citenamefont
  {Lichtenstein},\ and\ \citenamefont {Stepanov}}]{DGW}%
  \BibitemOpen
  \bibfield  {author} {\bibinfo {author} {\bibfnamefont {Nagamalleswararao}\
  \bibnamefont {Dasari}}, \bibinfo {author} {\bibfnamefont {Hugo U.~R.}\
  \bibnamefont {Strand}}, \bibinfo {author} {\bibfnamefont {Martin}\
  \bibnamefont {Eckstein}}, \bibinfo {author} {\bibfnamefont {Alexander~I.}\
  \bibnamefont {Lichtenstein}}, \ and\ \bibinfo {author} {\bibfnamefont
  {Evgeny~A.}\ \bibnamefont {Stepanov}},\ }\bibfield  {title} {\enquote
  {\bibinfo {title} {{Electron-magnon dynamics triggered by an ultrashort laser
  pulse: A real-time dual $GW$ study}},}\ }\href {\doibase 10.1103/vglv-2rmv}
  {\bibfield  {journal} {\bibinfo  {journal} {Phys. Rev. B}\ }\textbf {\bibinfo
  {volume} {111}},\ \bibinfo {pages} {235129} (\bibinfo {year}
  {2025})}\BibitemShut {NoStop}%
\bibitem [{\citenamefont {Tsuji}\ \emph {et~al.}(2009)\citenamefont {Tsuji},
  \citenamefont {Oka},\ and\ \citenamefont {Aoki}}]{Tsuji}%
  \BibitemOpen
  \bibfield  {author} {\bibinfo {author} {\bibfnamefont {Naoto}\ \bibnamefont
  {Tsuji}}, \bibinfo {author} {\bibfnamefont {Takashi}\ \bibnamefont {Oka}}, \
  and\ \bibinfo {author} {\bibfnamefont {Hideo}\ \bibnamefont {Aoki}},\
  }\bibfield  {title} {\enquote {\bibinfo {title} {Nonequilibrium steady state
  of photoexcited correlated electrons in the presence of dissipation},}\
  }\href {\doibase 10.1103/PhysRevLett.103.047403} {\bibfield  {journal}
  {\bibinfo  {journal} {Phys. Rev. Lett.}\ }\textbf {\bibinfo {volume} {103}},\
  \bibinfo {pages} {047403} (\bibinfo {year} {2009})}\BibitemShut {NoStop}%
\bibitem [{\citenamefont {Eckstein}\ and\ \citenamefont
  {Kollar}(2008)}]{Martin2008}%
  \BibitemOpen
  \bibfield  {author} {\bibinfo {author} {\bibfnamefont {Martin}\ \bibnamefont
  {Eckstein}}\ and\ \bibinfo {author} {\bibfnamefont {Marcus}\ \bibnamefont
  {Kollar}},\ }\bibfield  {title} {\enquote {\bibinfo {title} {Theory of
  time-resolved optical spectroscopy on correlated electron systems},}\ }\href
  {\doibase 10.1103/PhysRevB.78.205119} {\bibfield  {journal} {\bibinfo
  {journal} {Phys. Rev. B}\ }\textbf {\bibinfo {volume} {78}},\ \bibinfo
  {pages} {205119} (\bibinfo {year} {2008})}\BibitemShut {NoStop}%
\bibitem [{\citenamefont {Shao}\ \emph {et~al.}(2016)\citenamefont {Shao},
  \citenamefont {Tohyama}, \citenamefont {Luo},\ and\ \citenamefont
  {Lu}}]{Shao}%
  \BibitemOpen
  \bibfield  {author} {\bibinfo {author} {\bibfnamefont {Can}\ \bibnamefont
  {Shao}}, \bibinfo {author} {\bibfnamefont {Takami}\ \bibnamefont {Tohyama}},
  \bibinfo {author} {\bibfnamefont {Hong-Gang}\ \bibnamefont {Luo}}, \ and\
  \bibinfo {author} {\bibfnamefont {Hantao}\ \bibnamefont {Lu}},\ }\bibfield
  {title} {\enquote {\bibinfo {title} {Numerical method to compute optical
  conductivity based on pump-probe simulations},}\ }\href {\doibase
  10.1103/PhysRevB.93.195144} {\bibfield  {journal} {\bibinfo  {journal} {Phys.
  Rev. B}\ }\textbf {\bibinfo {volume} {93}},\ \bibinfo {pages} {195144}
  (\bibinfo {year} {2016})}\BibitemShut {NoStop}%
\bibitem [{\citenamefont {Dasari}\ \emph {et~al.}(2020)\citenamefont {Dasari},
  \citenamefont {Li}, \citenamefont {Werner},\ and\ \citenamefont
  {Eckstein}}]{Dasari}%
  \BibitemOpen
  \bibfield  {author} {\bibinfo {author} {\bibfnamefont {Nagamalleswararao}\
  \bibnamefont {Dasari}}, \bibinfo {author} {\bibfnamefont {Jiajun}\
  \bibnamefont {Li}}, \bibinfo {author} {\bibfnamefont {Philipp}\ \bibnamefont
  {Werner}}, \ and\ \bibinfo {author} {\bibfnamefont {Martin}\ \bibnamefont
  {Eckstein}},\ }\bibfield  {title} {\enquote {\bibinfo {title} {Revealing
  hund's multiplets in mott insulators under strong electric fields},}\ }\href
  {\doibase 10.1103/PhysRevB.101.161107} {\bibfield  {journal} {\bibinfo
  {journal} {Phys. Rev. B}\ }\textbf {\bibinfo {volume} {101}},\ \bibinfo
  {pages} {161107} (\bibinfo {year} {2020})}\BibitemShut {NoStop}%
\bibitem [{\citenamefont {Gole\ifmmode~\check{z}\else \v{z}\fi{}}\ \emph
  {et~al.}(2019)\citenamefont {Gole\ifmmode~\check{z}\else \v{z}\fi{}},
  \citenamefont {Eckstein},\ and\ \citenamefont
  {Werner}}]{PhysRevB.100.235117}%
  \BibitemOpen
  \bibfield  {author} {\bibinfo {author} {\bibfnamefont {Denis}\ \bibnamefont
  {Gole\ifmmode~\check{z}\else \v{z}\fi{}}}, \bibinfo {author} {\bibfnamefont
  {Martin}\ \bibnamefont {Eckstein}}, \ and\ \bibinfo {author} {\bibfnamefont
  {Philipp}\ \bibnamefont {Werner}},\ }\bibfield  {title} {\enquote {\bibinfo
  {title} {{Multiband nonequilibrium $GW+\text{EDMFT}$ formalism for correlated
  insulators}},}\ }\href {\doibase 10.1103/PhysRevB.100.235117} {\bibfield
  {journal} {\bibinfo  {journal} {Phys. Rev. B}\ }\textbf {\bibinfo {volume}
  {100}},\ \bibinfo {pages} {235117} (\bibinfo {year} {2019})}\BibitemShut
  {NoStop}%
\bibitem [{\citenamefont {Kova\ifmmode \check{c}\else
  \v{c}\fi{}evi\ifmmode~\acute{c}\else \'{c}\fi{}}\ \emph
  {et~al.}(2025)\citenamefont {Kova\ifmmode \check{c}\else
  \v{c}\fi{}evi\ifmmode~\acute{c}\else \'{c}\fi{}}, \citenamefont {Ferrero},\
  and\ \citenamefont {Vu\ifmmode \check{c}\else \v{c}\fi{}i\ifmmode
  \check{c}\else \v{c}\fi{}evi\ifmmode~\acute{c}\else \'{c}\fi{}}}]{Jaksa}%
  \BibitemOpen
  \bibfield  {author} {\bibinfo {author} {\bibfnamefont {Jeremija}\
  \bibnamefont {Kova\ifmmode \check{c}\else
  \v{c}\fi{}evi\ifmmode~\acute{c}\else \'{c}\fi{}}}, \bibinfo {author}
  {\bibfnamefont {Michel}\ \bibnamefont {Ferrero}}, \ and\ \bibinfo {author}
  {\bibfnamefont {Jak\ifmmode \check{s}\else~\v{s}\fi{}a}\ \bibnamefont
  {Vu\ifmmode \check{c}\else \v{c}\fi{}i\ifmmode \check{c}\else
  \v{c}\fi{}evi\ifmmode~\acute{c}\else \'{c}\fi{}}},\ }\bibfield  {title}
  {\enquote {\bibinfo {title} {{Toward Numerically Exact Computation of
  Conductivity in the Thermodynamic Limit of Interacting Lattice Models}},}\
  }\href {\doibase 10.1103/mm38-zttx} {\bibfield  {journal} {\bibinfo
  {journal} {Phys. Rev. Lett.}\ }\textbf {\bibinfo {volume} {135}},\ \bibinfo
  {pages} {016502} (\bibinfo {year} {2025})}\BibitemShut {NoStop}%
\bibitem [{\citenamefont {Eckstein}\ and\ \citenamefont
  {Werner}(2010)}]{PhysRevB.82.115115}%
  \BibitemOpen
  \bibfield  {author} {\bibinfo {author} {\bibfnamefont {Martin}\ \bibnamefont
  {Eckstein}}\ and\ \bibinfo {author} {\bibfnamefont {Philipp}\ \bibnamefont
  {Werner}},\ }\bibfield  {title} {\enquote {\bibinfo {title} {Nonequilibrium
  dynamical mean-field calculations based on the noncrossing approximation and
  its generalizations},}\ }\href {\doibase 10.1103/PhysRevB.82.115115}
  {\bibfield  {journal} {\bibinfo  {journal} {Phys. Rev. B}\ }\textbf {\bibinfo
  {volume} {82}},\ \bibinfo {pages} {115115} (\bibinfo {year}
  {2010})}\BibitemShut {NoStop}%
\bibitem [{\citenamefont {Schüler}\ \emph {et~al.}(2020)\citenamefont
  {Schüler}, \citenamefont {Golež}, \citenamefont {Murakami}, \citenamefont
  {Bittner}, \citenamefont {Herrmann}, \citenamefont {Strand}, \citenamefont
  {Werner},\ and\ \citenamefont {Eckstein}}]{SCHULER2020107484}%
  \BibitemOpen
  \bibfield  {author} {\bibinfo {author} {\bibfnamefont {Michael}\ \bibnamefont
  {Schüler}}, \bibinfo {author} {\bibfnamefont {Denis}\ \bibnamefont
  {Golež}}, \bibinfo {author} {\bibfnamefont {Yuta}\ \bibnamefont {Murakami}},
  \bibinfo {author} {\bibfnamefont {Nikolaj}\ \bibnamefont {Bittner}}, \bibinfo
  {author} {\bibfnamefont {Andreas}\ \bibnamefont {Herrmann}}, \bibinfo
  {author} {\bibfnamefont {Hugo~U.R.}\ \bibnamefont {Strand}}, \bibinfo
  {author} {\bibfnamefont {Philipp}\ \bibnamefont {Werner}}, \ and\ \bibinfo
  {author} {\bibfnamefont {Martin}\ \bibnamefont {Eckstein}},\ }\bibfield
  {title} {\enquote {\bibinfo {title} {{NESSi: The Non-Equilibrium Systems
  Simulation package}},}\ }\href {\doibase
  https://doi.org/10.1016/j.cpc.2020.107484} {\bibfield  {journal} {\bibinfo
  {journal} {Comput. Phys. Commun.}\ }\textbf {\bibinfo {volume} {257}},\
  \bibinfo {pages} {107484} (\bibinfo {year} {2020})}\BibitemShut {NoStop}%
\end{thebibliography}%

\end{document}


\title{Supplemental Material\\[0.3cm]
Nonlocal Correlation Effects in dc and Optical Conductivity of the Hubbard Model} 

\author{Nagamalleswararao Dasari}
\affiliation{Institut f{\"u}r Theoretische Physik, Universit{\"a}t Hamburg, Notkestra{\ss}e   9 , 22607 Hamburg, Germany}

\author{Hugo U. R. Strand}
\affiliation{School of Science and Technology, Orebro University, SE-70182 Obrebo, Sweden}

\author{Martin Eckstein}
\affiliation{Institut f{\"u}r Theoretische Physik, Universit{\"a}t Hamburg, Notkestra{\ss}e   9 , 22607 Hamburg, Germany}
\affiliation{The Hamburg Centre for Ultrafast Imaging, Luruper Chaussee 149, 22761 Hamburg, Germany}

\author{Alexander I. Lichtenstein}
\affiliation{Institut f{\"u}r Theoretische Physik, Universit{\"a}t Hamburg, Notkestra{\ss}e   9 , 22607 Hamburg, Germany}
\affiliation{European X-Ray Free-Electron Laser Facility, Holzkoppel 4, 22869 Schenefeld, Germany}
\affiliation{The Hamburg Centre for Ultrafast Imaging, Luruper Chaussee 149, 22761 Hamburg, Germany}

\author{Evgeny A. Stepanov}
\affiliation{CPHT, CNRS, \'Ecole polytechnique, Institut Polytechnique de Paris, 91120 Palaiseau, France}
\affiliation{Coll\`ege de France, 11 place Marcelin Berthelot, 75005 Paris, France}

\maketitle

\onecolumngrid
\section {optical conductivity}

The current operator is defined by the relation $J(r)$=$-c\delta H/\delta A(r)$. Here, $c$ is the velocity of light, $H$ is the Hubbard Hamiltonian, and $A(r)$ is the vector potential. In the Peierls approximation, the electric current in the long wave-length limit is given by ${\langle j(t) \rangle = \langle \frac{1}{V} \int d^d r J(r) e^{iqr}\rangle_{q\rightarrow 0}}$ and $V$ is the number of ${\bf k}$-points in the Brillouin zone. In the momentum representation 
\begin{equation}
\langle j_{\alpha}(t) \rangle= \frac{-ie}{V}\sum_{{\bf k},\sigma} v^{\alpha}_{\bf k}(t) G^{<}_{{\bf k}\sigma}(t,t),
\end{equation}
where the current vertex ${v^{\alpha}_{\bf k}(t) = \frac{1}{\hbar} \partial_{k_{\alpha}} {\varepsilon}_{{\bf k}-\frac{e}{\hbar c} {\bf A}(t)}}$. 
In 2d, the electronic band dispersion within Peierl's approximation has the following form: ${\varepsilon_{k-\frac{e}{\hbar c} A(t)} = -2J*[cos(k_{x}-\frac{e}{\hbar c} A(t))+cos(k_{y}-\frac{e}{\hbar c} A(t))]}$.
Since we are interested in the linear current response to the weak probe field, we define the current-current correlation function called susceptibility 
\begin{align}
{\chi_{\alpha \beta}(t,t') = \frac{\partial j_{\alpha}(t)}{\partial A_{\beta}(t')}\Big|_{A=0}},
\label{eq:chi}
\end{align} 
where $G^{<}_{\bf k}$ is the lesser component of the electronic Green's function. In the chosen gauge ${E(t)=-\partial_{t} A(t)}$, the susceptibilty $\chi_{\alpha \beta}(t,t')$ is related to the optical conductivity $\sigma_{\alpha \beta}(t,t')$ by
\begin{equation}
\sigma_{\alpha \beta}(t,t') = -c \int^{\infty}_{t'} \chi_{\alpha \beta} (t,\bar{t}) d \bar{t}.
\end{equation}
In equilibrium, the conductivity \(\sigma_{\alpha \beta}(t,t')\) depends solely on the time difference and thus the frequency-dependent conductivity $\sigma_{\alpha \beta}(\omega)$ can be obtained through a straightforward Fourier transform
\begin{align}
\text{Re} \left[\sigma_{\alpha \beta}(\omega)\right] = \int^{t_{max}}_0 \sigma_{\alpha \beta}(t - t') e^{i\omega (t - t')} \, d(t - t'), 
\end{align}
where $t_{max}$ is maximum simulated time. The susceptibility can be calculated by taking the derivative in Eq.~\eqref{eq:chi}, where the vector potential appears in both the velocity and Green's function. 
It gives a diagrammatic and paramagnetic contribution to the susceptibility,
\begin{equation}
 \chi_{\alpha \beta} (t,t') = \chi_{\alpha \beta}^{\rm dia} (t,t') + \chi_{\alpha \beta}^{\rm pm}(t,t'),
\end{equation}
with
\begin{equation}
\chi_{\alpha \beta}^{\rm dia} (t,t') = \frac{-ie}{V} \sum_{{\bf k},\sigma} \frac{\partial v^{\alpha}_{{\bf k}}(t)}{\partial A_{\beta}(t')} G^{<}_{{\bf k}\sigma}(t,t)~~~\text{and}~~~
\chi_{\alpha \beta}^{\rm pm} (t,t') = \frac{-ie}{V} \sum_{{\bf k},\sigma} v^{\alpha}_{{\bf k}}(t) \frac{\partial G^{<}_{{\bf k}\sigma}(t,t)}{\partial A_{\beta}(t')}.
\end{equation}
The diamagnetic contribution is further simplified and is given by
\begin{equation}
\chi_{\alpha \beta}^{\rm dia} (t,t') = \frac{-i\chi_0}{\hbar V} \sum_{{\bf k},\sigma} \partial_{k_\alpha}  \partial_{k_\beta} \varepsilon_{{\bf k}} (t) G^{<}_{{\bf k}\sigma}(t,t) \delta(t-t'),
\end{equation}
where the prefactor ${\chi_0 = \frac{e^2}{\hbar c}}$. The paramagnetic contribution is found from the variation of the lattice Dyson equation 
\begin{align}
\frac{\partial G_{{\bf k}\sigma}(t_1,t_2)}{\partial A_{\beta}(t)} 
=  - \iint \{dt'\} \,
G_{{\bf k}\sigma}(t_1,t'_1) \frac{\partial G^{-1}_{{\bf k}\sigma}(t'_1,t'_2)}{\partial A_{\beta}(t)}G_{{\bf k}\sigma}(t'_2,t_2)  ,
\end{align}
\begin{align}
\frac{\partial G^{-1}_{{\bf k}\sigma}(t_1,t_2)}{\partial A_{\beta}(t)} 
= \frac{\partial \left[G^{-1}_{0,{\bf k}\sigma}(t_1,t_2) - \Sigma_{{\bf k}\sigma}(t_1,t_2) \right]}{\partial A_{\beta}(t)}
= - \delta(t_1,t_2) \delta(t_1,t) \frac{e}{\hbar c} v^{\beta}_{\bf k}(t) - \frac{\partial \Sigma_{{\bf k}\sigma}(t_1,t_2)}{\partial A_{\beta}(t)}.
\end{align}

\begin{figure}[t!]
\includegraphics[width=0.6\linewidth]{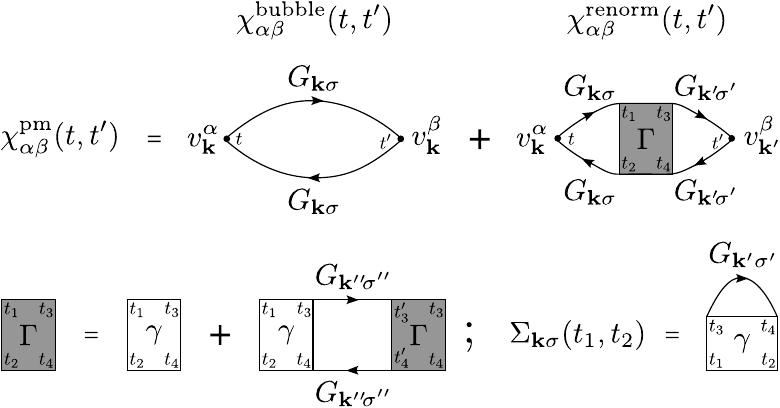}
\caption{Diagrammatic representation for the paramagnetic susceptibility ${\chi^{\rm pm}_{\alpha\beta}}$, vertex function ${\Gamma^{\sigma\sigma'}_{\bf k,k'}}$, and self-energy $\Sigma_{{\bf k}\sigma}$.} 
\label{fig:diagrams}
\end{figure}

\noindent
Considering the self-energy in the form: 
\begin{align}
\Sigma_{\bf k\sigma}(t_1,t_2) = i\iint dt_3\,dt_4 \sum_{\bf k',\sigma'} \gamma^{\sigma\sigma'}_{{\bf kk'}}(t_1,t_2,t_3,t_4)G_{{\bf k'}\sigma'}(t_3,t_4)
\label{eq:Sigma_SM}
\end{align}
and assuming that the vertex function does not change upon applying a small field, gives in the linear response approximation:
\begin{align}
\frac{\partial \Sigma_{{\bf k}\sigma}(t_1,t_2)}{\partial A_{\beta}(t)} = i\iint dt_3\,dt_4 \sum_{\bf k',\sigma'} \gamma^{\sigma\sigma'}_{{\bf kk'}}(t_1,t_2,t_3,t_4) \frac{\partial G_{{\bf k'}\sigma'}(t_3,t_4)}{\partial A_{\beta}(t)}.
\end{align}
Note that considering the self-energy in the exact Schwinger–Dyson form immediately identifies $\gamma$ as the exact two-particle irreducible vertex in the particle-hole channel.

Collecting all terms gives the final result for the paramagnetic susceptibility
${
\chi^{\rm pm}_{\alpha \beta}(t,t') = \chi^{\rm bubble}_{\alpha \beta}(t,t') + \chi^{\rm renorm}_{\alpha \beta}(t,t')
}$,
where the first, so called, ``bubble'' term
\begin{align}
\chi^{\rm bubble}_{\alpha \beta}(t,t') = \frac{-i\chi_0}{V}\sum_{{\bf k},\sigma} v^{\alpha}_{\bf k}(t) G_{{\bf k}\sigma}(t,t')G_{{\bf k}\sigma}(t',t) v^{\beta}_{\bf k}(t')
\label{eq:bubble_SM}
\end{align}
is expressed only through the Green's functions and velocities, and the second, renormalized, term
\begin{align}
&\chi^{\rm renorm}_{\alpha \beta}(t,t') = \iint \{dt_{i}\}\sum_{{\bf k,k'},\sigma,\sigma'} v^{\alpha}_{\bf k}(t) X^{0}_{{\bf k}\sigma}(t,t,t_1,t_2) \Gamma^{\sigma\sigma'}_{{\bf k,k'}}(t_1,t_2,t_3,t_4) X^{0}_{{\bf k'}\sigma'}(t_3,t_4,t',t') v^{\beta}_{\bf k'}(t')
\label{eq:renorm_SM}
\end{align}
additionally contains all possible multi-electron scattering processes through the correction $\Gamma$.
Importantly, the vertex correction $\Gamma$ is directly related to the scattering processes considered in the self-energy~\eqref{eq:Sigma_SM} via $\gamma$:
\begin{align}
\Gamma^{\sigma\sigma'}_{{\bf k,k'}}(t_1,t_2,t_3,t_4) = \gamma^{\sigma\sigma'}_{{\bf k,k'}}(t_1,t_2,t_3,t_4)
+ \iint \{dt'\} \sum_{\bf k'',\sigma''} \gamma^{\sigma\sigma''}_{{\bf k,k''}}(t_1,t_2,t'_1,t'_2) X^{0}_{{\bf k''}\sigma''}(t'_1,t'_2,t'_3,t'_4)\Gamma^{\sigma''\sigma'}_{{\bf k'',k'}}(t'_3,t'_4,t_3,t_4),
\end{align}
where 
\begin{align}
X^{0}_{{\bf k}\sigma}(t_1,t_2,t_3,t_4) = i G_{{\bf k}\sigma}(t_1,t_3)G_{{\bf k}\sigma}(t_4,t_2).
\end{align}
In particular, in \mbox{D-TRILEX}~\cite{PhysRevB.100.205115, PhysRevB.103.245123, 10.21468/SciPostPhys.13.2.036} and ${D\text{-}GW}$~\cite{DGW} the $\gamma$ vertex in the self-energy contains the charge and spin fluctuations via the renormalized charge and spin interaction $W$, which results in the vertex correction $\Gamma$ that is related to a multi-electron scattering on these charge and spin fluctuations.
Importantly, if the $\Gamma$ vertex is local (momentum-independent), the renormalised contribution to susceptibility~\eqref{eq:renorm_SM} disappears, because the integrand becomes odd in momenta ${\bf k}$ and ${\bf k'}$.
The diagrammatic representations of the susceptibility, vertex function, and self-energy are shown in Fig.~\ref{fig:diagrams}.

In this work, we focus on calculating the optical conductivity, referred to as ``bubble,'' which involves the susceptibility based on both diamagnetic and paramagnetic terms. Traditional ``full'' calculations require a renormalized susceptibility, necessitating a solution to the Bethe-Salpeter equation; however, this approach is computationally intensive. To address this challenge, we adopt a different strategy. We measure the linear current response of the system by applying a weak probe field and then compute the optical conductivity using the equation:
${\sigma(\omega) = J_{\rm pr}(\omega)/E_{\rm pr}(\omega)}$.
In this formulation, we apply a time-dependent probe field of the form 
${E_{\rm pr}(t) = E_0 \frac{(t - t_d)}{t^2_c} e^{-(t - t_d)^2 / (2t^2_c)}}$, where $t_c$ represents the pulse duration and $t_d$ is the pulse delay. The probe amplitude $E_0$ is kept sufficiently weak so that $\sigma(\omega)$ can be measured in the linear response limit, ensuring it does not depend on the amplitude of the probe. The probe current $J_{\rm pr}(\omega)$ measured under this weak pulse captures all vertex corrections associated with multi-particle scattering processes, which is a significant advantage of real-time methods~\cite{Tsuji, Martin2008, Shao, Dasari, PhysRevB.100.235117, Jaksa}.

\section{Details of numerical calculations}

The ${D\text{-}GW}$ method enables a self-consistent approach to local correlations within dynamical mean-field theory (DMFT), accounting for spatial charge and spin fluctuations simultaneously within a diagrammatic framework. The numerical implementation of the method is discussed in detail in Ref.~\cite{DGW}. 
In this Letter, we address the DMFT problem on an L-shaped Keldysh contour using a real-time strong-coupling impurity solver called the non-crossing approximation (NCA) implemented through NESSi~\cite{PhysRevB.82.115115, SCHULER2020107484}. Calculations are performed for the periodic lattice of size ${20\times20}$. 
We choose the Matsubara grid with ${\Delta\tau=0.00333}$ and ${\Delta\tau=0.0001}$ values for high and low temperature calculations, respectively. For metals, we select a real-time grid with ${\Delta t = 0.012}$, while for Mott insulators, we use ${\Delta t = 0.010}$. This ensures that dynamical quantities converge appropriately with the grid size. To resolve the quasi-particle peak and pseudo-gap features in the optical conductivity, we perform numerical simulations up to ${t_{max} = 36}$. However, for Mott insulators, we can identify the gap features with a maximum simulation time of ${t_{max} = 20}$. The probe pulse was applied with an amplitude of ${E_0 = 0.01}$, a pulse duration of ${t_c = 0.08}$, and a pulse delay of ${t_d = 0.35}$. 

\section{Strength of magnetic fluctuations}

\begin{figure}[b!]
\includegraphics[width=0.95\columnwidth]{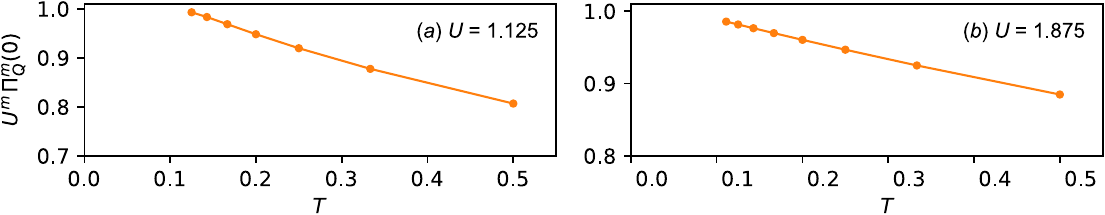}
\caption{The strength of magnetic fluctuations estimated by the product of the static (${\omega=0}$) polarization operator at the AFM wave vector ${Q=\{\pi,\pi\}}$, i.e. $\Pi^{m}_{\bf Q}(0)$, multiplied by the bare spin interaction $U^{m}$. The results are obtained using {\it D-GW} in the metallic [${U=1.125}$, panel (a)] and Mott-insulating [${U=1.875}$, panel (b)] regimes of the Hubbard model.}
\label{fig:correlation_s}
\end{figure}

The magnetic susceptibility $X^{m}_{\bf q}(\omega)$ can be expressed through the polarization operator $\Pi^{m}_{\bf q}(\omega)$ as: ${X^{m}_{\bf q}(\omega) = \Pi^{m}_{\bf q}(\omega) [1 - U^{m}\Pi^{m}_{\bf q}(\omega)]^{-1}}$, where ${U^{m}=-U/2}$ is the bare interaction in the spin channel. 
Therefore, the strength of magnetic fluctuations can be quantified by the value of the static (${\omega=0}$) polarization operator at the wave vector of the leading instability ${{\bf q}=Q}$ (AFM ${Q=\{\pi,\pi\}}$ in our case) multiplied by the bare spin interaction, i.e. ${U^{m}\Pi^{m}_{Q}(0)}$. 
If this quantity is zero, the system lacks spin fluctuations. 
On the contrary, if ${U^{m}\Pi^{m}_{Q}(0)}$ approaches unity, magnetic fluctuations become very strong. 
In this limit, the spin susceptibility diverges, signaling a phase transition to an ordered state.
The strength of AFM fluctuations estimated using {\it D-GW} in the metallic [${U=1.125}$, panel (a)] and Mott-insulating [${U=1.875}$, panel (b)] regimes of the Hubbard model is shown in Fig.~\ref{fig:correlation_s}.
We find that pronounced magnetic fluctuations extend to temperatures significantly above the AFM transition.

\section{Conductivity scans}

\begin{figure}[b!]
\includegraphics[width=0.9\columnwidth]{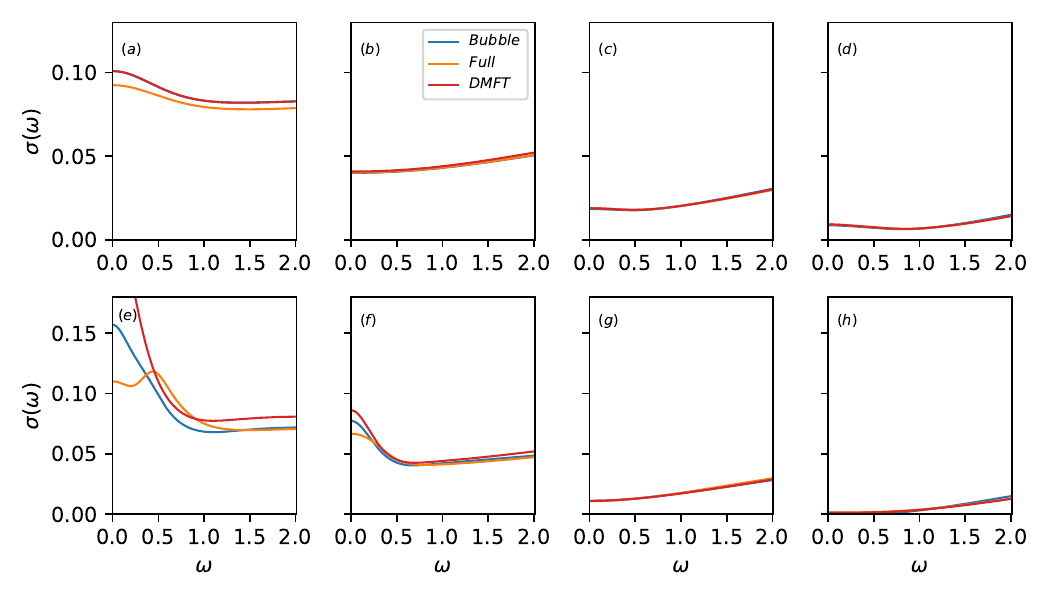}
\caption{The low-frequency part of the optical conductivity obtained at ${T=0.083}$ (top row) and ${T=0.036}$ (bottom row) for different values of the interaction $U$ using the bubble ${D\text{-}GW}$ (blue curve), full ${D\text{-}GW}$ (orange curve) and DMFT (red curve).
At ${U = 1.31}$ (a, e) the system lies in the metallic regime at both temperatures. 
${U = 1.525}$ (b, f) corresponds to the critical interaction for the Mott transition at the higher temperature, while at the lower temperature the system is still metallic. 
${U = 1.7}$ (c, g) corresponds to the critical interaction for the Mott transition at the lower temperature.
At ${U = 1.875}$ (d, h) the system resides in the Mott insulating state at both temperatures.}   
\label{fig:optical_scan}
\end{figure}

In Fig.~\ref{fig:optical_scan}, we show fixed-temperature scans of the optical conductivity obtained for different values of the interaction using the bubble ${D\text{-}GW}$ (blue curve), full ${D\text{-}GW}$ (orange curve) and DMFT (red curve) approaches. 
The difference between the bubble and full ${D\text{-}GW}$ methods corresponds to the contribution of vertex corrections to the conductivity, while the difference between the bubble ${D\text{-}GW}$ and DMFT results reflects the modification of the electronic spectral function by non-local correlations.
As expected, we find that at high temperatures, where non-local correlation effects on the spectral function are weak, the bubble ${D\text{-}GW}$ and DMFT results for the low-frequency part of the conductivity are nearly indistinguishable.
However, at low temperatures, these two results differ significantly in the metallic case, indicating that non-local spin fluctuations substantially renormalize the electronic spectral function, leading to a strong suppression of the low-frequency conductivity.
Upon entering the Mott insulating regime, which is predominantly governed by local correlations, this difference disappears.
Furthermore, by comparing the bubble and full ${D\text{-}GW}$ methods, we find that in the metallic regime (panels a, e, f), vertex corrections are significant, but their contribution decreases with increasing interaction strength, eventually vanishing at the critical interaction for the Mott transition (panels b, g).
In the Mott insulating regime (panels c, d, h), the bubble and full ${D\text{-}GW}$ approaches yield identical results for the low-frequency part of the conductivity, indicating that vertex corrections vanish from the DC conductivity in the finite temperature Mott phase. 

\section{Effect of the non-local Coulomb interaction on spectra and optics}

The interplay of spatial charge and spin fluctuations in quantum systems can significantly influence electronic spectral features and, consequently, their optical response. At the simplest level, spatial charge density wave fluctuations can be incorporated by introducing a short-range Coulomb interaction $V$ between the densities on the neighboring lattice sites ${\langle ij \rangle}$. The resulting Hamiltonian of the extended Hubbard model reads:
\begin{equation}
\mathcal{H} = 
-t\sum_{\langle ij \rangle, \sigma} c^{\dagger}_{i\sigma}c^{\phantom{\dagger}}_{j\sigma}
+ U \sum_i n^{\phantom{\dagger}}_{i\uparrow} n^{\phantom{\dagger}}_{i\downarrow} 
+ V\sum_{\langle ij \rangle,\sigma\sigma'} n^{\phantom{\dagger}}_{i\sigma} n^{\phantom{\dagger}}_{j\sigma'}, 
\label{eq:ex-model}
\end{equation}
In the following, we present the {\it D-GW} results obtained for different values of $V$ in the metallic and insulating regimes.

\subsection{Metal (${U = 1.125}$, ${T = 0.036}$)}

\begin{figure}[b!]
\includegraphics[width=0.95\linewidth]{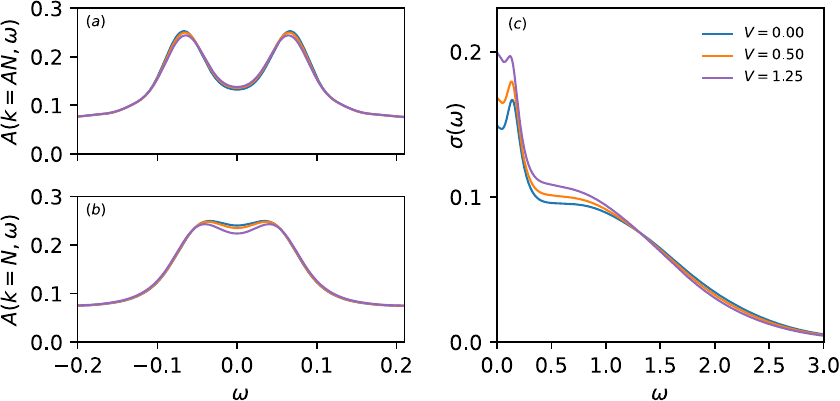}
\caption{The electronic spectral function ${A({\bf k},\omega)}$ obtained using {\it D-GW} in the metallic regime (${U = 1.125}$, ${T = 0.036}$) for different strengths of the non-local Coulomb interaction $V$ at the AN (a) and N (b) points of the BZ. The corresponding optical conductivity $\sigma(\omega)$ is shown in panel (c).
\label{fig:V-U45-effect}}
\end{figure}

In the metallic regime, in the absence of the non-local interaction (${V=0}$), a pseudogap dichotomy between the antinodal (AN) and nodal (N) points of the Brillouin zone (BZ) is observed in the single-particle spectral function, as shown in (a) and (b) panels of Fig.~\ref{fig:V-U45-effect}. 
While the charge fluctuations induced by the non-local Coulomb interaction in metallic systems are expected to suppress pseudogap features arising from magnetic fluctuations, the mechanism by which the system loses its pseudogap dichotomy remains unclear. 
Our {\it D-GW} results show that increasing $V$ fills the pseudogap at the AN point through a redistribution of spectral weight from the gap edges [Fig.~\ref{fig:V-U45-effect}\,(a)].
In contrast, increasing $V$ leads to a reduction of the spectral weight at the Fermi level at the N point [Fig.~\ref{fig:V-U45-effect}\,(b)]. 
This non-monotonic behavior at high-symmetry points in the BZ indicates an anisotropic influence of charge fluctuations on quasiparticle excitations in metallic systems.
Momentum-resolved spectra alone do not reveal whether the total gap in the system is enhanced or suppressed, in contrast to momentum-averaged probes.
Indeed, the optical conductivity, being a momentum-integrated response, exhibits an enhancement of the low-energy Drude weight [Fig.~\ref{fig:V-U45-effect}\,(c)], effectively filling the pseudogap and signaling an overall increase in metallicity.

\subsection{Mott-insulator (${U = 1.875}$,  ${T = 0.027}$)}

\begin{figure}[t!]
\includegraphics[width=0.65\columnwidth]{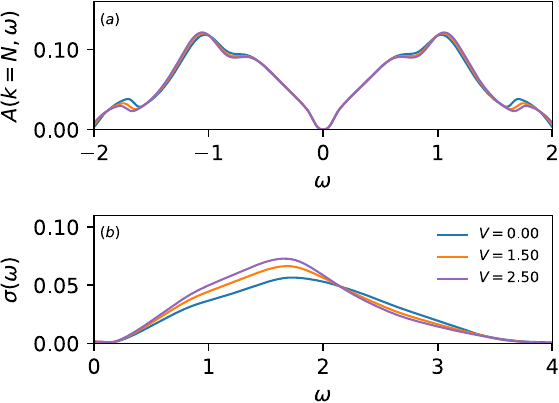}
\caption{The electronic spectral function ${A({\bf k},\omega)}$ obtained using {\it D-GW} in the Mott insulating regime (${U = 1.875}$,  ${T = 0.027}$) for different strengths of the non-local Coulomb interaction $V$ at the N (a) points of the BZ. 
The corresponding optical conductivity $\sigma(\omega)$ is shown in panel (b).}
\label{fig:V-U75-effect}
\end{figure}

The {\it D-GW} study of the Mott insulating regime at ${V=0}$ reveals a characteristic three-peak structure of each Hubbard band [Fig.~\ref{fig:V-U75-effect}\,(a)], which is a hallmark of strong magnetic fluctuations as discussed in the main text. 
These features are momentum independent across the BZ, reflecting the dominant role of local correlations in the Mott phase. 
Upon introducing the non-local Coulomb interaction, the overall structure of the electronic spectral function remains essentially unchanged, demonstrating the robustness of magnetic fluctuations in the Mott insulating phase.
We find that, with increasing $V$, the splitting of the Hubbard-band peaks deepens: the two peaks around ${\omega\approx1}$ move further apart, with the main (central) peak shifting to higher energies and the neighboring peak to lower energies [Fig.~\ref{fig:V-U75-effect}\,(a)]. 
Consistent with this picture, the optical conductivity displays a three-peak structure, with the peaks becoming more pronounced upon increasing $V$.

\bibliography{Ref}